\documentclass[]{jfm}

\usepackage{graphicx}
\usepackage{newtxtext}
\usepackage{newtxmath}
\usepackage{natbib}
\usepackage{hyperref}
\hypersetup{
    colorlinks = true,
    urlcolor   = blue,
    citecolor  = black,
}

\newcommand{\RomanNumeralCaps}[1]
\linenumbers
\usepackage{graphbox}

\newcommand{\dd}{\:\mathrm{d}}
\newcommand{\retau}{ \Rey_{\tau} }
\newcommand{\paren}[1]{\left( #1 \right)}
\newcommand{\bracket}[1]{\left[ #1 \right]}

\newcommand{\abs}[1]{\left| #1 \right|}

\usepackage{array}
\newcolumntype{C}[1]{>{\centering\let\newline\\\arraybackslash\hspace{0pt}}m{#1}}
\usepackage{multirow,multicol}
\usepackage{hhline}
\usepackage[table]{xcolor}
\definecolor{Grey}{HTML}{A9A9A9}
\definecolor{Blue}{HTML}{005ce6}
\definecolor{Red}{HTML}{ff3300}
\definecolor{Green}{HTML}{79ff4d}
\newcommand{\bm}[1]{\boldsymbol{#1}}

\newcommand{\real}[1]{\text{Re} \left\{ #1\right\} }
\newcommand{\innerproduct}[2]{\left\langle #1, \ #2 \right\rangle_y}

\newcommand{\expect}[1]{\mathbb{E}\left\{#1\right\}}
\newcommand{\e}[1]{\ensuremath{\times 10^{#1}}}

\title{Spatio-temporal characterization of nonlinear forcing and response in turbulent channel flow}

\author{Yuting Huang\aff{1}
  \corresp{\email{yhuang1@caltech.edu}},
  Simon S. Toedtli\aff{2},
  Gregory P. Chini\aff{3}
 \and 
 Beverley J. McKeon\aff{4}}

\affiliation{
	\aff{1}Mechanical and Civil Engineering, California Institute of Technology, Pasadena, CA 91125, USA
	\aff{2}National Center for Atmospheric Research, Boulder, CO 80301, USA
    \aff{3}Program in Integrated Applied Mathematics \& Department of Mechanical Engineering, University of New Hampshire, Durham, NH 03824, USA
	\aff{4}Department of Mechanical Engineering \& Center for Turbulence Research, Stanford University, Stanford, CA 94305, USA}

\begin{document}
\maketitle

\begin{abstract}
The quadratic convection term in the incompressible Navier-Stokes equations is considered as a nonlinear forcing to the linear resolvent operator, and it is studied in the Fourier domain through the analysis of interactions between triadically compatible wavenumber-frequency triplets. A framework to quantify the triadic contributions to the forcing and response by each pair of triplets is developed and applied to data from direct numerical simulations of a turbulent channel at $\retau \approx 550$. The linear resolvent operator is incorporated to provide the missing link from energy transfer between modes to the effect on the spectral turbulent kinetic energy. The coefficients highlight the importance of interactions involving large-scale structures, providing a natural connection to the modeling assumptions in quasi-linear (QL) and generalized quasi-linear (GQL) analyses. Specifically, it is revealed that the QL and GQL reductions efficiently capture important triadic interactions in the flow, especially when including of a small number of wavenumbers into the GQL large-scale base flow. Additionally, spatio-temporal analyses of the triadic contributions to a single mode representative of the near-wall cycle demonstrate the spatio-temporal nature of the triadic interactions and the effect of the resolvent operator, which selectively amplifies certain forcing profiles. The tools presented are expected to be useful for improving modeling of the nonlinearity, especially in QL, GQL, and resolvent analyses, and understanding the amplitude modulation mechanism relating large-scale fluctuations to the modulation of near-wall structures.
\end{abstract}

\begin{keywords}
Turbulent boundary layers, Turbulence modelling
\end{keywords}


\section{Introduction}
The beauty and complexity of turbulent flows arise in large part from the nonlinear convective terms in the Navier-Stokes equations (NSE). In many situations, the nonlinear terms remain an essential and challenging part of our understanding of turbulence. The most important role of this nonlinearity is the transfer of energy between the vast range of scales in turbulence~\citep{Jimenez_2012}, and recent experimental studies on turbulence control by \citet{Marusic_2021} demonstrated the importance of nonlinearity in achieving net power savings.  Under a classical Reynolds decomposition, i.e. the definition of turbulent fluctuations relative to a temporally- or spatio-temporally-averaged mean field, the quadratic nonlinearities in the incompressible NSE manifest themselves as a convolution of triadically compatible, i.e. resonant, interactions in the Fourier domain, linking a dyad of interacting scales to nonlinearity at a third scale. 

These nonlinear interactions and the associated spectral transfer have been studied in the spatial domain in the recent literature by, e.g. \cite{Cheung14}, \cite{Cho_Hwang_Choi_2018} and \cite{ding2024modetomodenonlinearenergytransfer}, and in the context of spatial and spectral fluxes by, e.g. \cite{Marati04}. The coherence between resonant spatial (or temporal) scales can also be studied through the skewness of the velocity and amplitude modulation of the small scales by large ones, e.g. \cite{marusic2010}, both of which can be expressed as a measure of relative phase between modes \citep{Duvvuri_McKeon_2015}. Similarly, \citet{Schmidt_2020} proposed the bispectral mode decomposition to study coherence in the velocity signals among spatial triads and analyzed the interacting frequency components using maxima in the mode bispectrum. 

The constraints on the resonance condition, i.e. which scales can interact, become stricter when spatio-temporal interactions are considered \citep{mckeon2017}. \cite{Barthel_2022} extended the observations of \cite{Schmid_Henningson_2001} concerning energy transfer in individual triads.

\citet{Karban_NL_23} have investigated the key triads underpinning minimal Couette flow. In earlier work \citep{huang2023spatiotemporalcharacterizationnonlinearforcing}, we documented the contributions of individual triads to the full nonlinear forcing in turbulent channel flow by quantifying the projection of individual dyad interactions in the wavenumber-frequency domain onto the forcing at the resonant scale, and characterizing the dominant interactions that arose.  Recently, \cite{yeung2024revealingstructuresymmetrynonlinearity} have exploited a similar approach in the so-called triadic orthogonal decomposition (TOD).

Most of these previous studies on triadic interactions have focused on the energy transfer between scales rather than linking the energy transfer to the spectral turbulence kinetic energy (TKE) or the velocity response attributable to the nonlinear forcing. Resolvent analysis \citep{mckeon2010} has emerged as a tool to identify spatio-temporal basis sets for the nonlinearity, which are treated as input forcing to the linear NSE (resolvent) operator and are preferentially amplified by the linear system giving rise to the state (velocity) response. In this framework, the nonlinearity has been treated crudely as a broadband input to analyze the properties of the linear operator by a range of authors or with more sophistication, e.g. as a colored forcing \citep{Zare_Jovanovic_Georgiou_2017}, which can be attributed to nonlinear modal interactions, to address the turbulence closure problem~\citep{mckeon2017}. 

Also relevant to the current work are quasi-linear (QL) models, in which the resolved nonlinear interactions are restricted to those either involving or resulting in the zero streamwise (streamwise constant) wavenumber modes \cite[e.g.][]{Farrell_Ioannou_2007}. Formally correct in the limit of separation of scales, the approach rests on the admitted interactions capturing the key elements of the nonlinearity, while the remaining unresolved interactions are neglected or approximated with a suitable model~\citep{Gayme_McKeon_Papachristodoulou_Bamieh_Doyle_2010, Farrell_Ioannou_2012}. Self-sustaining simulations can be achieved with flow features that resemble those obtained from direct numerical simulations (DNS), thus providing a cost-efficient model alternative to DNS of the full NSE~\citep{Thomas_Farrell_Ioannou_Gayme_2015, Farrell_Ioannou_Jimenez_Constantinou_Lozano_Nikolaidis_2016}. Generalized quasi-linear (GQL) analysis allows for resolving nonlinear interactions involving increasing numbers of harmonics of the streamwise fundamental wavelength associated with the domain, enabling spectrally non-local streamwise energy transfers \citep{Marston_Chini_Tobias_2016}; the complexity of the simulations increases as the filter separating large and small scales moves to smaller scale, converging to DNS in the limit of no filter. \citet{Farrell_Ioannou_2007, Marston_Chini_Tobias_2016} demonstrated the success of GQL on zonal jets in atmospheric turbulence and \citet{Kellam_2019, Hernandez_Yang_Hwang_2022a, Hernandez_Yang_Hwang_2022b} demonstrated the success of GQL for a turbulent channel flow. 
Finally, the selection of $\Lambda$ in GQL, the streamwise cutoff wavenumber between the large and small scales, currently largely relies on trials and comparisons with the baseline direct numerical simulations (DNS) or large eddy simulations (LES), and could benefit from a quantitative analysis of the important triadic interactions. To our knowledge, the importance of the resolved nonlinear interactions relative to the unresolved (neglected or modeled) ones in QL and GQL has not been fully quantified or explained, which is another motivation for the present study.

In this work, we aim to include the linear resolvent operator into the analysis of nonlinear energy transfer, to quantitatively characterize the spatio-temporal nature of the triadic interactions and their influence on the resulting velocity response in turbulent channel flow. Understanding the contribution of individual triads to overall forcing and response within the resolvent framework as well as within the spatio-temporal model of the energy cascade is the primary objective of this work. We begin by describing in Section 2 the space-time formulation for nonlinear interactions, the action of resolvent operator in exciting velocity response from the nonlinear forcing and the connection between this approach and other related observations. After validating the DNS approach in Section 3, we describe the contributions of individual scale interactions in exciting a near-wall mode in Section 4. Dominant interactions in the full channel domain are described in Section 5. Some implications for quasi-linear modeling are given in Section 6. Conclusions are given in Section 7.

\section{Formulation}\label{sec:formulation}
We consider an incompressible, fully-developed turbulent channel flow, with the streamwise, wall-normal and spanwise coordinates given by $x$, $y$, $z$, and the corresponding velocity components denoted by $u$, $v$, $w$. Vectors $\bm{x} = [x,y,z]$ and $\bm{u} = [u, v, w]$ are the spatial coordinate and velocity vectors, respectively. Unless otherwise specified, normalization of the coordinates and velocity components is performed using the outer scales: channel half height $h$ and channel centerline velocity $U_{CL}$.

For this fully developed channel flow with homogeneous streamwise and spanwise directions, the flow field is decomposed into the spatio-temporal mean profile $\overline{U}(y)$, averaged in $x$, $z$, $t$, and the perturbations $\bm{u}(x,y,z,t)$ relative to the spatio-temporal mean. The perturbation equations can then be obtained by subtracting the mean equations from the full NSE:
\begin{align}
	\nabla \cdot \bm{u} &= 0,\\
	\frac{\p \bm{u}}{\p t} + (\bm{u}\cdot\nabla) \overline{U} + (\overline{U}\cdot\nabla) \bm{u} &= - \nabla p + \frac{1}{\Rey} \nabla^2 \bm{u} + \bm{f},
\end{align}
where $p$ is the pressure fluctuations and $\bm{f}$ is the nonlinear forcing defined in physical space as
\begin{equation}
	\bm{f}(\bm{x},t) =  - \bm{u}(\bm{x},t) \cdot \nabla \bm{u}(\bm{x},t) + \overline{\bm{u}(\bm{x},t) \cdot \nabla \bm{u}(\bm{x},t)}. \label{eq:f}
\end{equation}
The nonlinear forcing is a result of grouping all terms that are nonlinear with respect to the perturbations stemming from the nonlinear convection term in the NSE. It may be treated crudely as a broadband input to analyze the properties of the linear operator or with more sophistication such as a data-driven forcing~\citep{Towne2020} or an eddy viscosity to address the turbulence closure problem~\citep{hwang2010linear}. 

A Fourier decomposition is employed in the homogeneous directions of $x$, $z$, and $t$:
\begin{equation}
	\bm{u}(\bm{x},t) = \iiint_{-\infty}^{\infty} \bm{{u}}(\bm{k},y) e^{i(k_x x+ k_z z - \omega t)} d k_x d k_z d \omega ,
\end{equation}
where we have introduced a wavenumber-frequency triplet $\bm{k} = [k_x, k_z, \omega]$. Here $k_x$, $k_z$ are the streamwise, spanwise wavenumbers, and $\omega$ is the temporal frequency. 

The Fourier-transformed perturbation equations are then written in an input-output form, where the nonlinear term $\bm{{f}} = [f_x, f_y, f_z]^T$ is considered as an input forcing to the resolvent operator $\mathcal{H}_p(\bm{k},y)$:
\begin{equation}
	\begin{bmatrix} \bm{{u}}(\bm{k},y) \\ {p}(\bm{k},y) \end{bmatrix} = \mathcal{H}_p(\bm{k},y) ~ \bm{{f}}(\bm{k},y). \label{eq:resolvent_nse}
\end{equation}
$\mathcal{H}_p(\bm{k},y)$ is the $4\times 3$ primitive form resolvent operator that maps the 3 forcing components to the 3 velocity components and pressure. Here, we focus on the $3\times 3$ operator $\mathcal{H}(\bm{k},y)$ governing the velocity response, where $\mathcal{H}_{ij}(\bm{k},y) = \mathcal{H}_{p_{ij}}(\bm{k},y)$ for $i, j = 1, 2,3$.

\subsection{Nonlinear forcing in the Navier-Stokes equations}\label{ch2:nl}
The nonlinear (quadratic) terms $\bm{f}$ are defined in physical space in equation~\ref{eq:f} through a point-wise multiplication, while in (discrete) Fourier space the nonlinear forcing at a wavenumber-frequency triplet $\bm{k}_3$ can be written in terms of a convolution of the velocity fields and velocity gradients at $\bm{k}_1$ and $\bm{k}_2$:
\begin{equation}
	\bm{f}(\bm{k}_3,y) = - \sum_{ \bm{k}_{1}+\bm{k}_{2} = \bm{k}_3}  \bm{u}(\bm{k}_1,y) \cdot \nabla  \bm{u}(\bm{k}_2,y). \label{eq:fconv}
\end{equation}
The requirement of $\bm{k}_1 + \bm{k}_2 = \bm{k}_3$ is the triadic compatibility or resonance constraint, a result of the quadratic nature of the nonlinearity.

In the resolvent formulation, the Fourier-transformed NSE are written in an input-output form, where $\bm{f}(\bm{k},y)$ is considered an input forcing to the resolvent operator $\mathcal{H}(\bm{k},y)$ as shown in equation~\eqref{eq:resolvent_nse}. It can be seen that the linear operator does not modify the scale of the input, such that each wavenumber-frequency triplet operates independently from all others. On the other hand, equation~\eqref{eq:fconv} shows that the nonlinear forcing is responsible for the coupling of different scales, and therefore the distribution of energy among scales.

\begin{figure}
	\centering
	\includegraphics[width=0.8\linewidth]{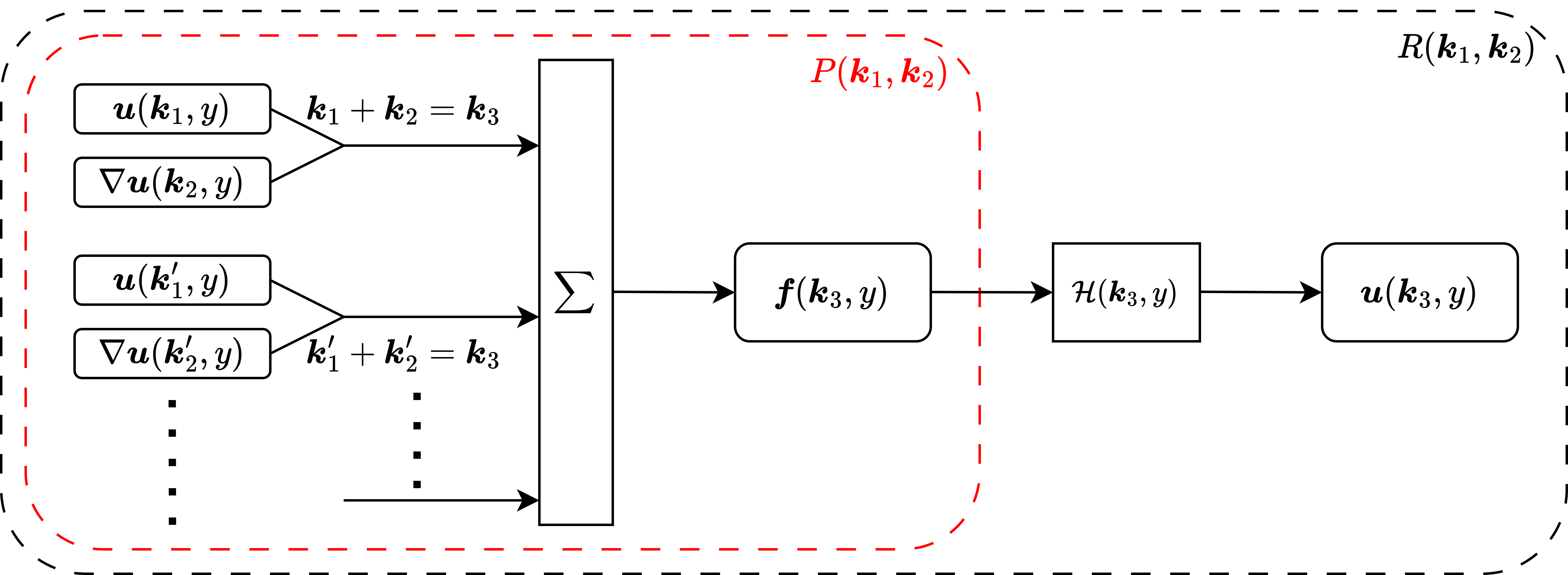}
	\caption{Diagram for the triadic interactions. The velocity and velocity gradient at $\bm{k}_1$ and $\bm{k}_2$ interact nonlinearly to generate part of the forcing at $\bm{k}_3 = \bm{k}_1 + \bm{k}_2$. The full forcing is a convolution sum of all pairs of $\bm{k}_1$ and $\bm{k}_2$ that are triadically compatible with $\bm{k}_3$, which forces the resolvent operator to generate the response. The red inner box contains the triadic contributions to the nonlinear forcing, studied using the coefficients $P(\bm{k}_{1}, \bm{k}_{2})$, and the black outer box contains the triadic contributions to the response, studied using the coefficients $R(\bm{k}_{1}, \bm{k}_{2})$.}
	\label{fig:triadicinteraction2} 
\end{figure}

The triadic interactions are visually depicted in Figure~\ref{fig:triadicinteraction2}. First, the velocity fields at $\bm{k}_1$ nonlinearly interact with the velocity gradients at $\bm{k}_2$, generating part of the forcing at $\bm{k}_3= \bm{k}_1 + \bm{k}_2$. This triadic contribution to the forcing is studied using the forcing coefficients $P(\bm{k}_{1}, \bm{k}_{2})$ defined later in section~\ref{subsec:forcing}. The forcing at $\bm{k}_3$ is then passed through the linear resolvent operator to generate the velocity response at the same wavenumber-frequency triplet $\bm{k}_3$. This triadic contribution to the velocity response, which involves both the nonlinear convolution and the linear resolvent operator, is studied using the response coefficients $R(\bm{k}_{1}, \bm{k}_{2})$ defined later in section~\ref{subsec:response}.

\subsection{From time to frequency} \label{ch4:fourier}
To perform the spatio-temporal analyses, the time domain data needs to be transformed into the Fourier domain. For this purpose, the Welch method is applied where the temporal snapshots are segemented, with the Hann window function applied to each temporal segment before taking the temporal Fourier transform. Finally, the results are averaged across all temporal segments.
To analyze the effect of the window function, similar to \citet{Nogueira_Morra_Martini_Cavalieri_Henningson_2021} and \citet{Morra_Nogueira_Cavalieri_Henningson_2021}, the momentum equation is multiplied by the window function $w(t)$, and the spatial dimensions are temporarily ignored as they do not play a role in this analysis of the temporal window function:
\begin{equation}
w(t) \frac{\p}{\p t} \bm{{u}}(t) + w(t)\mathcal{L}\bm{{u}}(t) = w(t) \bm{{f}}(t).
\end{equation}
In this equation, $\mathcal{L}$ is the linear part of the Navier-Stokes operator, and the equation can then be rewritten as
\begin{equation}
	\frac{\p}{\p t} \bracket{w(t)\bm{{u}}(t)} + \mathcal{L} \bracket{w(t)\bm{{u}}(t)} - \bm{{u}}(t)\frac{d}{d t}w(t) = \bracket{w(t)\bm{{f}}(t)}.
\end{equation}
Performing the Fourier transform in time on the signals with the window function applied: $\bm{{u}}(\omega) = \mathcal{F}\bracket{w(t)\bm{{u}}(t)}, \bm{f}(\omega) = \mathcal{F}\bracket{w(t)\bm{{f}}(t)}$ and defining the spurious forcing: $\bm{{s}}(\omega) = \mathcal{F}\bracket{\bm{{u}}(t)\frac{d}{d t}w(t)}$, the above equation can be rewritten using the resolvent operator $\mathcal{H}$:
\begin{equation}
	\bm{{u}}(\omega) = \mathcal{H} \paren{\bm{f}(\omega) +\bm{{s}}(\omega)},
\end{equation}
which can be reorganized as
\begin{equation}
	\bm{\tilde{u}}(\omega) = \bm{{u}}(\omega) - \mathcal{H}\bm{{s}}(\omega)  = \mathcal{H} \bm{f}(\omega). \label{eq:utilde}
\end{equation}
$\bm{\tilde{u}}(\omega)$ are the velocity Fourier modes corrected for the spurious forcing due to the window function, which satisfies the input-output form of the resolvent analysis introduced in equation~\eqref{eq:resolvent_nse}.

Additionally, the forcing Fourier modes, $\bm{f}(\omega) = \mathcal{F}\bracket{w(t)\bm{{f}}(t)}$, can be computed in two equivalent ways: 
\begin{equation}
	\bm{f}(\omega) = \mathcal{F} \bracket{w(t) \bm{{u}}(t) \cdot \nabla \bm{{u}}(t)} =  \mathcal{F} \bracket{\sqrt{w(t)} \bm{{u}}(t)} * \mathcal{F} \bracket{\sqrt{w(t)} \nabla \bm{{u}}(t)},
\end{equation}
where $*$ denotes the convolution operator. The first method computes $\bm{{f}}(t)$ in physical time, then applies the window function and takes the Fourier transform, while the second method computes $\bm{f}(\omega)$ in the Fourier space through a convolution. Note that to ensure the equivalence between the two methods, the second method requires the use of the $\sqrt{w(t)}$ as the window function applied to the velocity and velocity gradients. For this work, the periodic Hann window function is utilized in the analyses, and the choice of the window function is not expected to impact the overall structure of the resulting coefficients. 



\subsection{Singular value decomposition of the resolvent operator}
The discrete resolvent operator $\mathcal{H}(\bm{k},y)$ is constructed using a compact finite difference in $y$ matching the numerical scheme of the DNS described later in section~\ref{sec:dns}. A singular value decomposition (SVD) can then be performed on the discrete operator:
\begin{equation}
	\mathcal{H}(\bm{k},y) = \sum_{q} \psi_q(\bm{k},y) \sigma_q(\bm{k}) \phi_q^*(\bm{k},y), \label{eq:svd}
\end{equation}
where $\psi_q$ are the singular response modes (henceforth referred to as resolvent modes), $\sigma_q$ are the (ordered) singular values, and $\phi_q$ are the singular forcing modes. Superscript $(\cdot)^*$ denotes a complex conjugate. Velocity Fourier modes can then be expressed as
\begin{equation} \label{eq:rankn}
	\bm{{u}}(\bm{k},y)  = \sum_{q} \chi_q(\bm{k}) \sigma_q(\bm{k}) \psi_q(\bm{k},y),
\end{equation}
where $\chi_q(\bm{k}) = \phi^*_q(\bm{k},y) \bm{{f}}(\bm{k},y)$ are the nonlinear weights obtained by projecting the nonlinear forcing (if known) onto the singular forcing modes.

\subsection{Energy transfer by the nonlinear forcing} \label{sec2:tke}
We start by defining the spectral turbulence kinetic energy (TKE) as $e(\bm{k},y) = \abs{u(\bm{k},y)}^2 + \abs{v(\bm{k},y)}^2 + \abs{w(\bm{k},y)}^2$, which is the energy of Fourier modes at given $\bm{k}$. An equation for the spectral TKE can be written as
\begin{align}
	& \underbrace{
		\real{ u^*(\bm{k},y) v(\bm{k},y) \overline{U}'(y) } \vphantom{\frac{d^2}{dy^2}} 
	}_{\text{Production}} 
	+ \underbrace{ 
		\frac{k^2 }{\Rey} e(\bm{k},y) + \frac{1}{\Rey} \frac{d}{dy}u_i^*(\bm{k},y) \frac{d}{dy} u_i(\bm{k},y)
	}_{\text{Viscous Dissipation}} \nonumber\\[0.5em]
	+& \underbrace{
		\real{\frac{d}{dy} \bracket{ v^*(\bm{k},y) p(\bm{k},y)}}
	}_{\text{Pressure Transport}}
	- \underbrace{
		\frac{1}{2} \frac{1}{\Rey} \frac{d^2}{dy^2} e(\bm{k},y) \vphantom{\real{\frac{d}{dy}}} 
	}_{\text{Viscous Transport}} 
	= \underbrace{
		\real{ u_i^*(\bm{k},y) f_i(\bm{k},y) \vphantom{\frac{d}{dy}} }   
	}_{\text{Turbulent Transport}}, \label{eq:stke}
\end{align}
where $\real{\cdot}$ indicates the real part, and the summation notation is used with the subscripts $i$. 

The turbulent transport term can be alternatively written as
\begin{align}
	 u_i^*(\bm{k},y) f_i(\bm{k},y) =& \ \bm{u}^*(\bm{k},y) \cdot \bm{f}(\bm{k},y) \nonumber \\[0.5em]
	  =& - \bm{u}^*(\bm{k},y) \cdot \sum_{\bm{k}_1 + \bm{k}_2 = \bm{k} } \bm{u}(\bm{k}_1,y) \cdot \nabla \bm{u}(\bm{k}_2,y), \label{eq:turbulenttransport}
\end{align}
which shows that the turbulent transport is a term that involves the nonlinear forcing, and therefore transports energy between different wavenumber-frequency triplets $\bm{k}$. Additionally, \citet{Schmid_Henningson_2001, Barthel_2022} have demonstrated that the nonlinear turbulent transport is energy conserving on a triad by triad basis. 

Alternatively, the resolvent formulation $\bm{u}(\bm{k},y) = \mathcal{H}(\bm{k},y) ~ \bm{{f}}(\bm{k},y)$ can be utilized to obtain
\begin{align}
	e(\bm{k},y) =&  \ \bm{u}^*(\bm{k},y) ~ \mathcal{H}(\bm{k},y) \bm{{f}}(\bm{k},y) \nonumber \\[0.5em]
	 =& -\bm{u}^*(\bm{k},y) ~ \mathcal{H}(\bm{k},y) \sum_{\bm{k}_1 + \bm{k}_2 = \bm{k} } \bm{u}(\bm{k}_1,y) \cdot \nabla \bm{u}(\bm{k}_2,y). \label{eq:stke2}
\end{align}
The formulation of equations~\eqref{eq:stke} and \eqref{eq:stke2}, both derived from the NSE, are mathematically equivalent, yet they have slightly different interpretations. Equation~\eqref{eq:stke} is an energy balance equation. The energy transported into or out of $\bm{k}$ through nonlinear interactions (the turbulent transport term), is balanced by 4 other mechanisms, and is therefore not directly correlated to the increase or decrease of the spectral TKE at this $\bm{k}$. Equation~\eqref{eq:stke2} on the other hand, considers the nonlinearity as a forcing that drives the turbulent perturbations and activates the linear mechanisms of production, pressure, and viscosity (all contained in the linear resolvent operator $\mathcal{H}$), therefore providing a direct link between the nonlinearity and the spectral TKE.

\subsection{Triadic contributions to the forcing} \label{subsec:forcing}
The contribution from the interaction between a pair of triplets $\bm{k}_1$ and $\bm{k}_2$ to the resulting forcing at $\bm{k}_3 = \bm{k}_1 + \bm{k}_2$, can be quantified through a forcing coefficient $P(\bm{k}_{1}, \bm{k}_{2})$:
\begin{equation}
	P(\bm{k}_{1}, \bm{k}_{2}) =   - \expect{ \innerproduct{ 
		\bm{f}(\bm{k}_1 + \bm{k}_2,y)
	}{
		\bm{u}(\bm{k}_1,y) \cdot \nabla \bm{u}(\bm{k}_2,y)
	} }, \label{eq:defp}
\end{equation}
which is the inner product between $-\bm{u}(\bm{k}_1,y) \cdot \nabla  \bm{u}(\bm{k}_2,y)$ and $\bm{f}(\bm{k}_3,y)=\bm{f}(\bm{k}_1 + \bm{k}_2,y)$. $\expect{\cdot}$ is the expected value operator, indicating an average over all temporal segments (or different realizations) and the weighted inner product $\innerproduct{a(y)}{b(y)}$ is defined as the integral over a certain $y$ range in the continuous domain and approximated in the discrete space using a weight matrix $W$ with the appropriate integration coefficients on the diagonal:
\begin{equation}
	\innerproduct{a(y)}{b(y)} = \int_y a^*(y) b(y) \dd y \approx a^* W b. \label{eq:innerproduct}
\end{equation}
In this work, three separate wall-normal ($y$) ranges will be used, loosely corresponding to the near-wall  $y^+ \in (0, 30)$, overlap $y^+ \in (30, 200)$, and wake regions $y^+ \in (200, 550)$, to study the wall-normal variations of these coefficients.

The intentionally un-normalized coefficients take the forcing magnitude into consideration, e.g. a large fractional contribution to a small magnitude forcing is treated as unimportant. Note that these coefficients, defined for DNS Fourier modes, differ from those, e.g. in \citet{mckeon2017}, which are defined for the interactions between resolvent modes. 

The results span the (large) parameter space spanned by $(k_x, k_z, \omega, y)$. To facilitate computation and visualization of $P(\bm{k}_{1}, \bm{k}_{2})$, we define $P_{k_x}$, $P_{k_z}$, and $P_{\omega}$ by summation in 4 of the 6 scale dimensions:
\begin{subequations}
\begin{align}
	P_{k_x}(k_{x1}, k_{x2}) &=  \sum_{k_{z1}}\sum_{k_{z2}}\sum_{\omega_1}\sum_{\omega_2}  P(\bm{k}_{1}, \bm{k}_{2}), \label{eq:defpkx}\\
	P_{k_z}(k_{z1}, k_{z2}) &=  \sum_{k_{x1}}\sum_{k_{x2}}\sum_{\omega_1}\sum_{\omega_2}  P(\bm{k}_{1}, \bm{k}_{2}), \label{eq:defpkz}\\
	P_{\omega}(\omega_1, \omega_2) &=  \sum_{k_{x1}}\sum_{k_{x2}}\sum_{k_{z1}}\sum_{k_{z2}}  P(\bm{k}_{1}, \bm{k}_{2}). \label{eq:defpom}
\end{align}
\label{eqs:p}
\end{subequations}
$P_{k_x}$ defined in equation~\eqref{eq:defpkx} describes the interaction in the streamwise direction between $k_{x1}$ and $k_{x2}$, summed over all possible $k_z$ and $\omega$ interactions. Similarly, $P_{k_z}(k_{z1}, k_{z2})$ describes the interaction between $k_{z1}$ and $k_{z2}$, and $P_{\omega}$ describes the interaction between $\omega_1$ and $\omega_2$.

Using the Hermitian symmetry of the velocity Fourier modes, it can be shown that the forcing coefficients are also Hermitian functions. However, they are asymmetric about their two arguments, due to the action of the velocity gradient tensor. The forcing coefficient satisfies:
\begin{equation}
	P(\bm{k}_{1}, \bm{k}_{2}) = P^*(-\bm{k}_{1}, -\bm{k}_{2})\neq P(\bm{k}_{2}, \bm{k}_{1}), \label{eq:hermitiansymmetry}
\end{equation}
with $P_{k_x}$, $P_{k_z}$, and $P_{\omega}$ satisfying the same property. We retain the two separate coefficients associated with each combination of $(\bm{k}_{1}, \bm{k}_{2})$ to maximize the information about the dominant interactions within each triad, which will be lost under a combined symmetric coefficient.

Utilizing equation~\eqref{eq:fconv}, we can obtain the summation property of the forcing coefficient:
\begin{equation}
	\sum_{ \bm{k}_1 } P(\bm{k}_{1}, \bm{k}_{3} - \bm{k}_{1}) = \expect{\innerproduct{ \bm{f}(\bm{k}_3,y) }{ \bm{f}(\bm{k}_3,y) }},
\end{equation} 
where the right-hand side of the equation is the spectral energy of the forcing at $\bm{k}_3$, a real positive quantity. As a result, the forcing coefficients can be interpreted as the triadic contributions to the forcing spectral energy. The positive and negative real parts of the coefficients represent energy injection and extraction respectively, while the complex parts of the coefficients cancel out in an integral sense due to the Hermitian symmetry described in equation~\eqref{eq:hermitiansymmetry}, providing an additional constraint on the ensemble of nonlinear interactions and resulting in a real positive sum. Similar properties can be obtained for $P_{k_x}$, $P_{k_z}$ and $P_{\omega}$:
\begin{subequations}
	\begin{align}
		\sum_{k_{x1}} P_{k_x}(k_{x1}, k_{x3}-k_{x1}) &= \sum_{k_{z3}}\sum_{\omega_3}  \expect{ \innerproduct{ \bm{f}(\bm{k}_3,y) }{ \bm{f}(\bm{k}_3,y) } },\\
		\sum_{k_{z1}} P_{k_z}(k_{z1}, k_{z3}-k_{z1}) &= \sum_{k_{x3}}\sum_{\omega_3}  \expect{ \innerproduct{ 	\bm{f}(\bm{k}_3,y) }{ \bm{f}(\bm{k}_3,y) } },\\
		\sum_{\omega_1} P_{\omega}(\omega_{1}, \omega_{3}-\omega_{1}) &= \sum_{k_{x3}}\sum_{k_{z3}}  \expect{ \innerproduct{ \bm{f}(\bm{k}_3,y) }{ \bm{f}(\bm{k}_3,y) } }. 
	\end{align}
\end{subequations}

\subsection{Triadic contributions to the velocity response} \label{subsec:response}
The forcing contributions at all $\bm{k}$s are not equally amplified. To study the triadic contribution from the interaction between a pair of triplets $\bm{k}_1$ and $\bm{k}_2$ to the resulting velocity response at $\bm{k}_3 = \bm{k}_1 + \bm{k}_2$, including the effect of the linear resolvent operator, we will pass $-\bm{u}(\bm{k}_1,y) \cdot \nabla  \bm{u}(\bm{k}_2,y)$ through $\mathcal{H}(\bm{k}_3,y)$ and then take the inner product with the resulting velocity response $\bm{\tilde{u}}(\bm{k}_3,y)$ to define the response coefficient:
\begin{equation}
	R(\bm{k}_{1}, \bm{k}_{2}) =  - \expect{ \innerproduct{ 
		\bm{\tilde{u}}(\bm{k}_1+\bm{k}_2,y)} { \mathcal{H}(\bm{k}_1+\bm{k}_2,y) \bracket{\bm{u}(\bm{k}_1,y) \cdot \nabla  \bm{u}(\bm{k}_2,y)}
	}}, \label{eq:defr}
\end{equation}
where $\bm{\tilde{u}}(\bm{k}_3,y)$, introduced previously in equation~\ref{eq:utilde}, are the velocity Fourier modes with a correction to remove the effect of the window function in the temporal Fourier transform. The coefficients are again intentionally un-normalized to take the response magnitude into consideration. 

Similar to the previous section, the 2-dimensional coefficients $R_{k_x}$, $R_{k_z}$, and $R_{\omega}$ are defined by summation in 4 of the 6 dimensions:
\begin{subequations}
\begin{align}
	R_{k_x}(k_{x1}, k_{x2}) &=  \sum_{k_{z1}}\sum_{k_{z2}}\sum_{\omega_1}\sum_{\omega_2}  R(\bm{k}_{1}, \bm{k}_{2}), \label{eq:defrkx}\\
	R_{k_z}(k_{z1}, k_{z2}) &=  \sum_{k_{x1}}\sum_{k_{x2}}\sum_{\omega_1}\sum_{\omega_2}  R(\bm{k}_{1}, \bm{k}_{2}), \label{eq:defrkz}\\
	R_{\omega}(\omega_1, \omega_2) &=  \sum_{k_{x1}}\sum_{k_{x2}}\sum_{k_{z1}}\sum_{k_{z2}}  R(\bm{k}_{1}, \bm{k}_{2}). \label{eq:defrom}
\end{align}
\label{eqs:R}
\end{subequations}
The Hermitian symmetry and the asymmetry about the two arguments are also satisfied by the response coefficient:
\begin{equation}
	R(\bm{k}_{1}, \bm{k}_{2}) = R^*(-\bm{k}_{1}, -\bm{k}_{2})\neq R(\bm{k}_{2}, \bm{k}_{1}), \label{eq:hermitiansymmetry2}
\end{equation}
with $R_{k_x}$, $R_{k_z}$, and $R_{\omega}$ satisfying the same property.

Utilizing equations~\eqref{eq:fconv} and \eqref{eq:utilde}, the summation properties of the response coefficient can also be obtained:
\begin{equation}
	\sum_{ \bm{k}_1 } R(\bm{k}_{1}, \bm{k}_{3} - \bm{k}_{1}) = \expect{ \innerproduct{ \bm{\tilde{u}}(\bm{k}_3,y) }{ \bm{\tilde{u}}(\bm{k}_3,y) } },
\end{equation} 
where the right-hand side of the equation is the spectral turbulent kinetic energy of the velocity response at $\bm{k}_3$, and is again a real positive quantity. As a result, the response coefficients can be interpreted as the triadic contributions to the spectral turbulent kinetic energy (TKE). The positive and negative real parts of the coefficients represent the injection and extraction of spectral TKE at a given wavelength or frequency, while the complex parts of the coefficients cancel out due to the Hermitian symmetry described in equation~\eqref{eq:hermitiansymmetry2}. Similar properties can be obtained for $R_{k_x}$, $R_{k_z}$, and $R_{\omega}$:
\begin{subequations}
	\begin{align}
		\sum_{k_{x1}} R_{k_x}(k_{x1}, k_{x3}-k_{x1}) &= \sum_{k_{z3}}\sum_{\omega_3}  \expect{ \innerproduct{ \bm{\tilde{u}}(\bm{k}_3,y) }{ \bm{\tilde{u}}(\bm{k}_3,y) } }, \label{eq:rkx_sum}\\
		\sum_{k_{z1}} R_{k_z}(k_{z1}, k_{z3}-k_{z1}) &= \sum_{k_{x3}}\sum_{\omega_3}  \expect{ \innerproduct{ \bm{\tilde{u}}(\bm{k}_3,y) }{ \bm{\tilde{u}}(\bm{k}_3,y) } }, \label{eq:rkz_sum}\\
		\sum_{\omega_1} R_{\omega}(\omega_{1}, \omega_{3}-\omega_{1}) &= \sum_{k_{x3}}\sum_{k_{z3}}  \expect{ \innerproduct{ \bm{\tilde{u}}(\bm{k}_3,y) }{ \bm{\tilde{u}}(\bm{k}_3,y) } }. \label{eq:rom_sum} 
	\end{align}
\end{subequations}

The calculation of these coefficients is computationally intensive in the wavenumber-frequency domain.  Thus, for efficiency, they are performed in the physical domain for all variables except the one parameterizing the coefficient.  For example, $P_{k_x}$ is calculated in $(z,t)$, 
\begin{align}
    &P_{k_x}(k_{x1}, k_{x2}) \nonumber\\
	&= -\frac{1}{{N_z N_t}} \sum_{z} \sum_{t} 
	 \expect{ \innerproduct{
		\bm{f}(k_{x1}+k_{x2},y,z,t)} { \bm{u}(k_{x1},y,z,t) \cdot \nabla\bm{u}(k_{x2},y,z,t) } }
\end{align}	
This form provides the alternative physical interpretations of the coefficients as quantifying the importance of interactions between $k_{x1}$ and $k_{x2}$ averaged over all spanwise locations and time instead of a summation over wavenumbers and frequencies in the Fourier domain. 

A similar procedure can be followed for $P_{k_z}$, $P_{\omega}$, $R_{k_x}$, $R_{k_z}$ and $R_\omega$, ensuring that the full 6 dimensional $P(\bm{k}_{1}, \bm{k}_{2})$ is never computed, reducing computation cost, memory and storage requirements. Further details may be found in \cite{Huang_2025}.

\subsection{Relation to triple correlation, bispectrum and trispectrum}
The forcing and response coefficients studied here may be related to the more well-known triple correlation and bispectrum. Following \citet{Lii_Rosenblatt_Atta_1976}, the three-point spatial triple correlation for three state variables $q_l$, $q_m$, and $q_n$ can be defined as:
\begin{equation}
	R_{lmn}(\bm{r},\bm{r'}) = \langle q_l(\bm{x})q_m(\bm{x}+\bm{r})q_n(\bm{x}+\bm{r'}) \rangle_{\bm{x}},
\end{equation}
where $\langle \cdot \rangle_{\bm{x}}$ represents a spatial average. Two triple spatial Fourier transforms of $R_{lmn}(\bm{r},\bm{r'})$ in $\bm{r}, \bm{r'}$ lead to the three-dimensional spatial bispectrum:
\begin{equation}
	B_{lmn}(\bm{\hat{k}}, \bm{\hat{k}'}) = q_l^*(\bm{\hat{k}}+\bm{\hat{k}'})q_m(\bm{\hat{k}})q_n(\bm{\hat{k}'}).
\end{equation}
The forcing coefficients proposed in this work in equation~\eqref{eq:defp} can be seen as a spatio-temporal extension of the bispectrum, considering the spatio-temporal wavenumbers, $\bm{k}$, that is suitable for the analysis of wall-bounded flows with the three homogeneous directions of $x$, $z$, and $t$. An additional difference is that the forcing coefficients in this work are specifically designed to study the contribution of an interacting pair on the resulting forcing, which to our knowledge has not been studied before. The three terms contributing to the interaction coefficients are the velocity $\bm{u}$, velocity gradient $\nabla\bm{u}$, and forcing $\bm{f}$ in contrast to previous studies that generally focused on the three terms being the same flow quantity: for example, $\p u/\p x$ in \citet{Lii_Rosenblatt_Atta_1976} to study the nonlinear energy transfer between scales and velocity $\bm{u}$ in \citet{Schmidt_2020}, where the bispectral mode decomposition is introduced to compute modes associated with triadic interactions through maximization of the integral bispectral density. 

Both the Bispectral Mode Decomposition (BMD) \citep{Schmidt_2020} and the Triadic Orthogonal Decomposition (TOD) \citep{yeung2024revealingstructuresymmetrynonlinearity} consider third-order statistics. BMD focuses on the two-point auto-bispectral density matrix: $\bm{B}(\bm{x}, \bm{x'}, \omega_1, \omega_2) =  \bm{u}(\bm{x},\omega_1) \bm{u}(\bm{x},\omega_2) \bm{u}^*(\bm{x}',\omega_1 + \omega_2)$, while TOD focuses on the two-point cross-bispectral covariance tensor, $\bm{S}(\bm{x}, \bm{x'}, \omega_1, \omega_2) =  \left[ \bm{u}(\bm{x},\omega_1) \cdot \nabla \right] \bm{u}(\bm{x},\omega_2) \bm{u}^*(\bm{x}',\omega_1 + \omega_2)$. Both methods have implied spatial cross-correlation and aim to extract optimal modes with respect to certain metrics. However, without consideration of the linear operator, both BMD and TOD  are directly related to the turbulent transport term in equation~\eqref{eq:stke}, and characterize the transfer of energy between different frequencies. This work, on the other hand, does not intend to extract optimal modes, but aims to further the understanding of the combined effect of the linear and nonlinear mechanisms. The response coefficients proposed in equation~\eqref{eq:defr} can be considered as a weighted version of the spatio-temporal extension to the bispectrum. This spatio-temporal extension considers spatial wavenumbers, which allows for the study of nonlinear triadic interactions between structures of different spatial sizes. More importantly, the response coefficients include the effect of the linear resolvent operator, acting as the weight matrix. This inclusion of the linear resolvent operator takes into account the effects of other terms in the spectral TKE equation~\eqref{eq:stke}, such as viscosity, pressure, and production by the mean shear. For example, energy transfer from one pair of $(\bm{k}_1,\bm{k}_2)$ to $\bm{k}_3 = \bm{k}_1 + \bm{k}_2$ with energy distributed mainly at a wall-normal location with strong mean shear could potentially increase the mode amplitude at this location and induce a stronger production, resulting in some energy amplification, therefore contributing more significantly to the spectral TKE at $\bm{k}_3$, while another pair transferring energy to the same $\bm{k}_3$ but with energy distributed mainly at another $y$ location with weak mean shear will not have this effect. The inclusion of the linear resolvent operator as a weight matrix accounts for these mechanisms, and therefore provides a new and more complete picture of the effects of the triadic energy transfer on the resulting spectral TKE. Additionally, the nonlinear forcing can be decomposed into irrotational and solenoidal parts using the Helmholtz decomposition \citep{Rosenberg_2018, Morra_Nogueira_Cavalieri_Henningson_2021}. The irrotational part of the forcing has no effect on the resulting velocity fields, and is naturally eliminated when multiplied by the resolvent operator. Therefore, the response coefficient defined in equation~\eqref{eq:defr} naturally removes the effect of the inactive irrotational forcing, which differs from both the forcing coefficients defined in equation~\eqref{eq:defp} and the bispectrum analyses.

\section{Data from direct numerical simulation} \label{sec:dns}
\begin{figure}
	\centering
	\includegraphics[width=380pt]{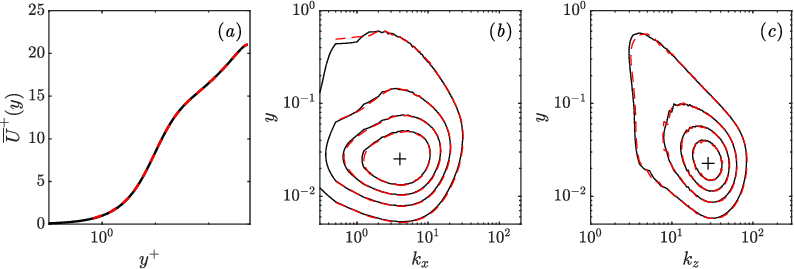}
	\caption{Comparison of (\textit{a}) the spatio-temporal mean in inner scales $\overline{U}^+(y^+)$, and (\textit{b}, \textit{c}) the temporal averaged pre-multiplied $uu$ power spectra between the results of \citet{Lee_Moser_2015} (black solid line) and the current DNS study (red dashed line). The $uu$ power spectra are plotted using contour lines at the same levels for both the results of \citet{Lee_Moser_2015} and the current study. The + markers in figure (\textit{b}) located at $k_x = 4$ and in figure (\textit{c}) at $k_z = 28$ mark the peak in the $uu$ power spectra, which are the representative wavenumbers for the near wall cycle.} 
	\label{fig:mean_spectra}
\end{figure}

The two sets of interaction coefficients described above are evaluated using data from a modified DNS code of \citet{Flores_Jimenez_2006}. The channel half height is denoted as $h$, the domain size is $4\pi h \times 2 \pi h$, and the friction Reynolds number $\retau = u_{\tau} h/\nu$ is approximately 551. The code uses a spectral discretization in the streamwise ($x$) and spanwise ($z$) directions, with the nonlinear terms computed in physical space with $2/3$ dealiasing, and a compact finite differences scheme in the wall-normal direction ($y$), with total number of points $N_y$. Quantities normalized with inner-units, using the viscous length scale $\delta_{\nu} = \nu / u_{\tau}$ and friction velocity $u_\tau$ are indicated with superscripts `$+$'. Otherwise, normalization is performed with channel center-line velocity $U_{cl}$ and channel half-height $h$. The simulation parameters are similar to previous full and QL turbulent channel simulations and are compared to studies of~\citet{Lee_Moser_2015} and \citet{Flores_Jimenez_2006} in table~\ref{tab:dnspara}. Although the simulation box size $L_x \times L_z = 4\pi h \times 2 \pi h$ is relatively small, it is a commonly utilized size in previous studies, especially for QL systems, and is less expensive for computation and data storage than a longer domain. The maximum wavenumbers retained by the DNS are $k_x = \pm 127.5$, $k_z = \pm 255$. The time stepping is performed using a third-order Runge-Kutta scheme with constant step sizes $\Delta t = 0.00185$, and sampled every 20 time steps with a sampling time of $\Delta t_s = 0.0369$, both normalized by $U_{cl}$ and $h$. The sample rate is selected to capture most of the energetic frequencies while keeping the data size manageable. A total of 6144 DNS snapshots are collected, with a total eddy turnover time of $T u_{\tau}/h=10.78$. Turbulence statistics are in good agreement with previous studies as shown in Figure~\ref{fig:mean_spectra} for the spatio-temporal mean profile and the temporal averaged $uu$ power spectra. All analyses performed on the DNS data set, including the resolvent analysis, utilize identical spatial grids and spatial differentiation schemes as the DNS, while in the temporal domain, a Fourier analysis is utilized instead of time stepping, the implementation of which is discussed previously in section~\ref{ch4:fourier}.

\begin{table}
	\centering 
	\begin{tabular}{cccccccccc} 
		&$\retau$&$L_x/h$&$L_z/h$&$\Delta x^+$&$\Delta z^+$&$\Delta y_{wall}^+$&$\Delta y_{cl}^+$&$N_y$&$T u_{\tau}/h$  \\    \hline 
		\cite{Lee_Moser_2015}     &  544   &$8\pi$ &$3\pi$ & 8.9        & 5.0        & 0.019        & 4.5          & 384 & 13.6           \\    
		\cite{Flores_Jimenez_2006}&  556   &$4\pi$ &$2\pi$ & 10.2       & --         & 0.8          & 7.0          & --  & --             \\  
		Current Study             &  551   &$4\pi$ &$2\pi$ & 9.0        &4.5         &0.69          & 6.0          & 272 & 10.78 \\
		\hline
	\end{tabular} 	
    \caption{Comparison of simulation parameters used by~\citet{Lee_Moser_2015}, \citet{Flores_Jimenez_2006} and the current study.}
	\label{tab:dnspara}
\end{table}

Although the selected sampling rate captures most of the energetic frequencies of the velocity fluctuations, it is insufficient to capture all the frequencies of the nonlinear forcing, due to the quadratic nature resulting in higher frequencies. To prevent aliasing, a temporal low pass filter is added into the DNS to remove the high-frequency content before down sampling. The filter is demonstrated in \cite{Huang_2025} to successfully prevent aliasing in the nonlinear forcing while introducing no phase distortion to the data. 

Utilizing the Welch method discussed in section~\ref{ch4:fourier}, the 6,144 DNS temporal snapshots are segmented into 5 segments of 2,048 snapshots each, with 50\% overlap. The Hann window function is then applied to each temporal segment before taking the temporal Fourier transform, and the frequencies within the resolved frequency range $\pm \omega_R = \pm 2\pi f_R = \pm 42.53$ are retained.

\begin{figure}
	\centering
	\includegraphics[width=377pt]{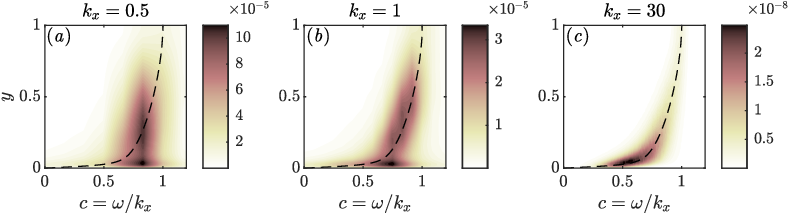}
	\caption{Contour plots of streamwise power spectra as a function of $k_x$, $y$, and the wavespeed $c = \omega/k_x$. Figures (\textit{a, b}) correspond to the large scales with $k_x = 0.5, 1$, and (\textit{c}) to a small scale with $k_x = 30$. The black dashed lines are the spatio-temporal averaged streamwise velocity profile $\overline{U}(y)$, which also marks the critical layer locations.}
	\label{fig:spec} 
\end{figure}

The window function and Fourier transform are applied to the streamwise velocity modes to compute the streamwise power spectra $\phi_{uu}(c,y;k_x)$, plotted as a function of $k_x$, $y$, and the wavespeed $c = \omega/k_x$ for two representative large scales $k_x = 0.5, 1$ in Figure~\ref{fig:spec}(\textit{a}) and (\textit{b}) and a representative small scale at $k_x = 30$ in Figure~\ref{fig:spec}(\textit{c}). The black dashed lines in each subplot show the streamwise mean velocity profile, $\overline{U}(y)$. It can be observed that the energy contained in the large scales is distributed in $y$, extending almost throughout the entire channel height, and predominantly located at high wavespeeds. On the other hand, the small scales (large $k_x$) have most of the energy located at smaller wavespeeds and have an energy distribution that is mostly localized near the wall and in a $y$ range that is centered around the local mean velocity.

Next, for verification of the temporal Fourier analysis, the following two quantities are computed utilizing the Welch method and the Hann window function described in section~\ref{ch4:fourier} for all wavenumber-frequency triplets $\bm{k}$:
\begin{eqnarray}
	V(\bm{k}) &=& \expect{ \innerproduct{ \bm{\tilde{u}}(\bm{k},y) } { \mathcal{H}(\bm{k},y) \bm{f}(\bm{k},y) } }, \label{eq:v}\\
	E_u(\bm{k}) &=& \expect{ \innerproduct{ \bm{\tilde{u}}(\bm{k},y) } { \bm{\tilde{u}}(\bm{k},y) } } , \label{eq:ek}
\end{eqnarray}
where $\expect{\cdot}$ is the expected value operator, indicating an average over all temporal segments. $E_u(\bm{k})$ is defined as the spectral turbulent kinetic energy of $\bm{\tilde{u}}(\bm{k})$ at a given wavenumber-frequency triplet $\bm{k}$. Utilizing equation~\eqref{eq:utilde}, the two quantities should be equal with $V(\bm{k})/E_u(\bm{k}) = 1$. A similar demonstration of this agreement between $\bm{\tilde{u}}$ and $\mathcal{H} \bm{f}$ is performed in~\citet{Morra_Nogueira_Cavalieri_Henningson_2021}, where the agreement is characterized using the power spectral density as a function of $y$ for selected modes. Here, we elect to demonstrate the agreement in a $y$-integrated sense for all computed wavenumber-frequency triplets.

\begin{figure}
	\centering
	\includegraphics[width=375pt]{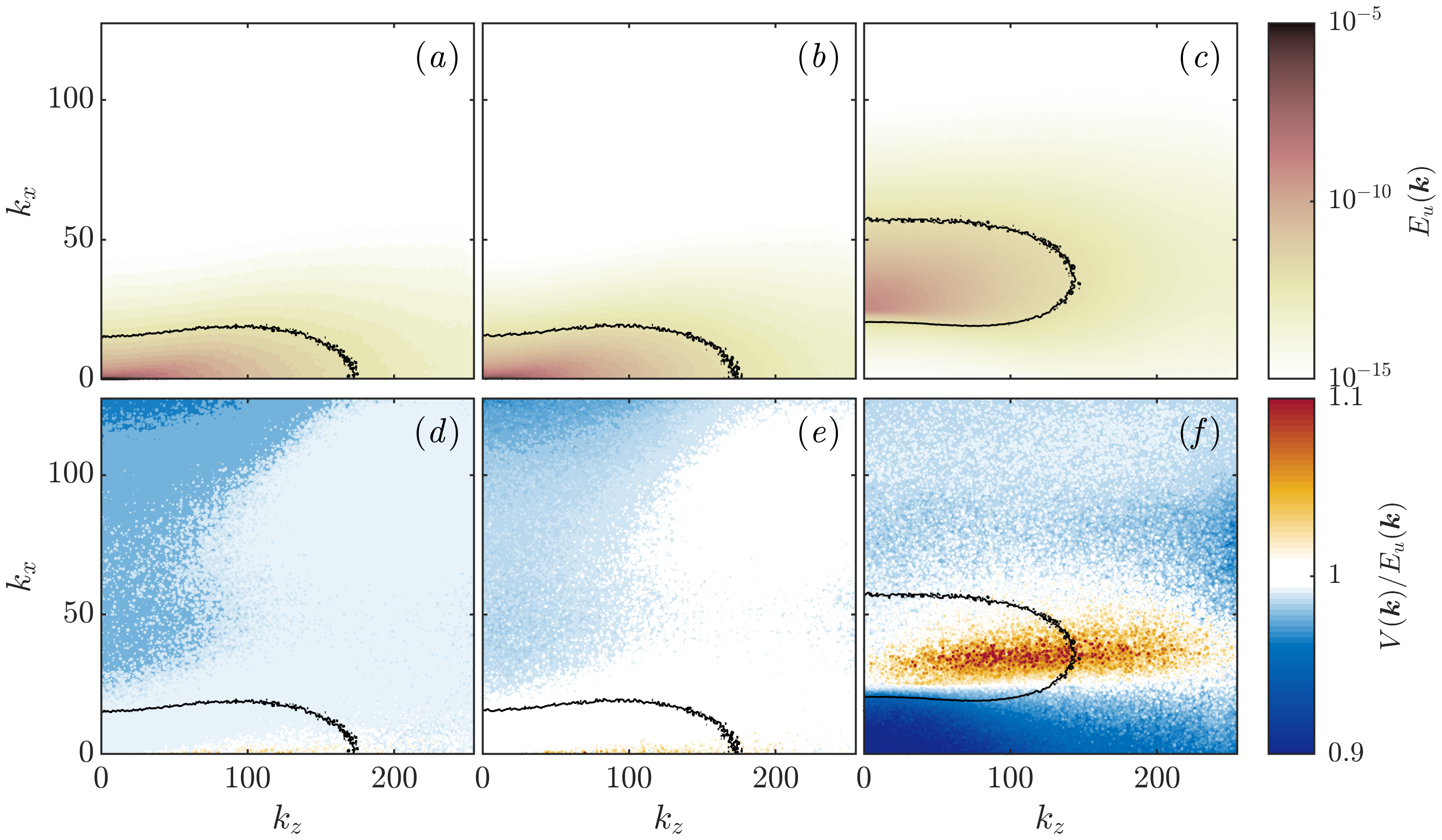}
	\caption{Contour plots of the comparison between $E_u(\bm{k})$ and $V(\bm{k})$. The top row (\textit{a - c}) are $E_u(\bm{k})$, the spectral turbulent kinetic energy of $\bm{\tilde{u}}$ with a log scale colorbar, and the bottom row (\textit{d - f}) are $V(\bm{k})/E_u(\bm{k})$ with a linear colorbar. The three columns are $\omega = 0, 0.166, 25.088$, and the black contour lines in all subplots are the energy level of $10^{-12}$.}
	\label{fig:v}
\end{figure}

In Figure~\ref{fig:v}, the quantities $E_u(\bm{k})$ and $V(\bm{k})/E_u(\bm{k})$ are plotted in the $k_x - k_z$ planes for two representative low frequencies $\omega = 0, 0.166$ and one high frequency $\omega= 25.088$ to examine the agreement between $V(\bm{k})$ and $E_u(\bm{k})$. The black contour lines in all figures denote the energy level of $10^{-12}$, which is more than 10 orders of magnitude weaker than the most energetic modes in the flow. In other words, the regions outside of the black contour lines are of less dynamic significance. From Figure~\ref{fig:v}(\textit{d}) and (\textit{e}), it can be observed that for the low frequencies, $V(\bm{k})$ and $E_u(\bm{k})$ agree very well with each other, especially in the high energy regions enclosed by the black contour lines. As the frequency increases, the agreement degrades. Although still in relatively good agreement in the high energy regions, the low energy regions start to show increasing errors. This behavior is expected, primarily for two reasons: first, as the frequency increases, the spectral TKE of the modes decreases, leading to higher sensitivity to numerical errors when normalizing by the spectral TKE in $V(\bm{k})/E_u(\bm{k})$. Additionally, high frequency modes are expected to receive more contributions from the nonlinear interactions involving high frequencies (further demonstrated in later sections), part of which are removed by the low pass filter for sampling, resulting in larger errors. However, these differences are not expected to affect the computation of either the forcing or response coefficients as the intentionally un-normalized coefficients are designed to receive small contributions from the low energy modes. Finally, although only the results averaged over all temporal segments are shown in Figure~\ref{fig:v}, the agreement for each temporal segment has been found to be essentially the same as for the averaged results.

\section{Triadic contributions to a single $\bm{k}_3$ representative of the near-wall cycle}\label{sec:traid_nwc}

We begin by analyzing the individual triadic interactions that contribute to excite a single Fourier mode at $\bm{k}_3$. The peaks in the pre-multiplied time-averaged streamwise power spectra in $k_x-y$ and $k_z-y$ planes in Figures~\ref{fig:mean_spectra}(\textit{b - c}) representative of the near-wall cycle, marked with + markers, correspond to wavenumbers $k_{x3} = 4$ and $k_{z3} = 28$. The most energetic frequency at these wavenumbers is $\omega_3 = 2.492$. The resulting mode of $\bm{k}_3 = [4, \  28, \ 2.492]$ has a wavespeed of $c_3 = \omega_3/k_{x3} = 0.62$, and corresponds to $\lambda_x^+  = 865$, $\lambda_z^+ = 124$, $\omega_3^+ = 0.0953$, and $c_3^+ = \omega_3^+/k_{x3}^+ = 13$ in inner scales.

\begin{figure}
	\centering
	\includegraphics[width=350pt]{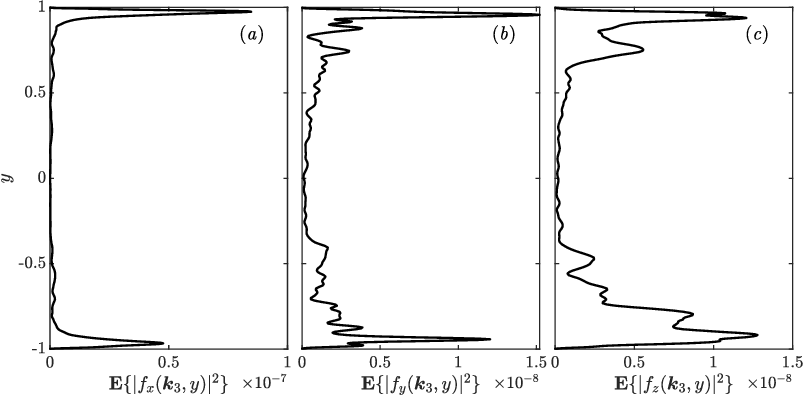}
	
	\vspace*{1em}
	
	\includegraphics[width=350pt]{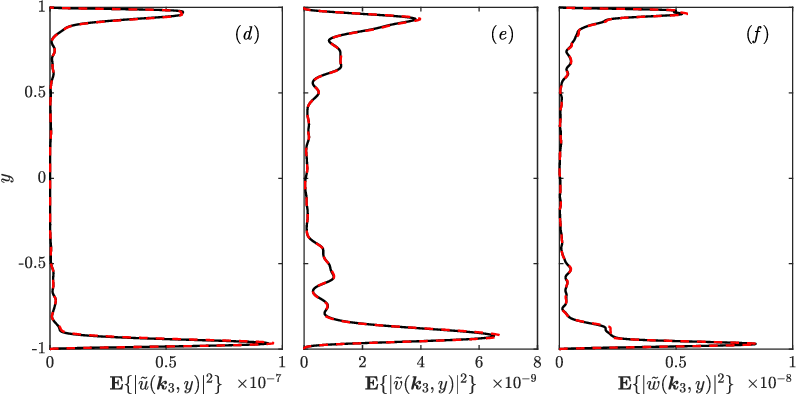}
	\caption{Spectral energy of Fourier modes for $\bm{k}_3 = \bm{k}_1+\bm{k}_2 = [4, \ 28, \ 2.492]$ averaged over all temporal segments. The top row (\textit{a - c}) are the three components of $\expect{ \abs{\bm{f}(\bm{k}_3,y)}^2 }$, the energy of the forcing Fourier modes. The  black solid lines in bottom row (\textit{d - f}) are the three components of $\expect{\abs{\bm{\tilde{u}}(\bm{k}_3, y)}^2}$, the energy of the velocity Fourier modes corrected for the effect of the window function, and the red dashed lines are for $\expect{\abs{\mathcal{H}(\bm{k}_3,y) \bm{f}(\bm{k}_3,y)}^2}$.} 
	\label{fig:k3fouriermode}
\end{figure}

The energy of the forcing and velocity Fourier modes are averaged across the different temporal segments, $\expect{\abs{\bm{f}(\bm{k}_3, y)}^2}$ and $\expect{\abs{\bm{\tilde{u}}(\bm{k}_3, y)}^2}$, to give the spectral energy as a function of $y$. This process of averaging over multiple temporal segments is the Welch's method to estimate the PSD \citep{Welch_1967}, as similarly employed in the works of \citet{Towne_Schmidt_Colonius_2018}, \citet{Nogueira_Morra_Martini_Cavalieri_Henningson_2021} and \citet{Morra_Nogueira_Cavalieri_Henningson_2021}. The energy of the forcing and velocity Fourier modes, $\expect{\abs{\bm{f}(\bm{k}_3, y)}^2}$ and $\expect{\abs{\bm{\tilde{u}}(\bm{k}_3, y)}^2}$, are plotted in Figure~\ref{fig:k3fouriermode} together with the TKE $\expect{\abs{\mathcal{H}(\bm{k}_3,y) \bm{f}(\bm{k}_3,y)}^2}$ for comparison. The agreement between $\bm{\tilde{u}}(\bm{k}_3, y)$ and  $\mathcal{H}(\bm{k}_3,y) \bm{f}(\bm{k}_3,y)$, first demonstrated in Figure~\ref{fig:v} in an $y$-integrated sense for all $\bm{k}_3$, can also be seen at the level of interactions contributing to this near-wall mode.  
Very close alignment can be observed between the black lines for $\expect{\abs{\bm{\tilde{u}}(\bm{k}_3, y)}^2}$ and red lines for $\expect{\abs{\mathcal{H}(\bm{k}_3,y) \bm{f}(\bm{k}_3,y)}^2}$ in the bottom row of Figure~\ref{fig:k3fouriermode}. All five individual temporal segments at this $\bm{k}_3$ exhibit similar levels of close agreement between $\bm{\tilde{u}}(\bm{k}_3, y)$ and $\mathcal{H}(\bm{k}_3,y) \bm{f}(\bm{k}_3,y)$ (not shown).


We introduce the notation for the individual contributions to the forcing and velocity Fourier modes by the interaction between $\bm{k}_1$ and $\bm{k}_2$:
\begin{subequations}
	\begin{eqnarray}
		\bm{f}(\bm{k}_1,\bm{k}_2,y) &=& - \bm{u}(\bm{k}_1,y) \cdot \nabla  \bm{u}(\bm{k}_2,y), 
	\label{eq:ftriad}\\[0.5em]
		\bm{\tilde{u}}(\bm{k}_1,\bm{k}_2,y) &=& \mathcal{H}(\bm{k}_1+\bm{k}_2,y) \bm{f}(\bm{k}_1,\bm{k}_2,y), 	\label{eq:utriad}
	\end{eqnarray}
\end{subequations}
and the energy of each term, averaged over all temporal segments, is defined as
\begin{subequations}
	\begin{eqnarray}
		E_f(\bm{k}_1, \bm{k}_2) 
		&=& \expect{\innerproduct{\bm{f}(\bm{k}_1,\bm{k}_2,y)}{\bm{f}(\bm{k}_1,\bm{k}_2,y)}}, \label{eq:ef}\\[0.5em]
		E_u(\bm{k}_1, \bm{k}_2) 
		&=& \expect{\innerproduct{\bm{\tilde{u}}(\bm{k}_1,\bm{k}_2,y)}{\bm{\tilde{u}}(\bm{k}_1,\bm{k}_2,y)}} , \label{eq:eu}
	\end{eqnarray}
\end{subequations}
with the inner product $\innerproduct{\cdot \ }{\cdot}$ defined previously in equation~\eqref{eq:innerproduct}. 

The convolution form of the nonlinear forcing, equation~\eqref{eq:fconv}, leads to the expressions for the full forcing and response at a given $\bm{k}_3$:
\begin{subequations}
	\begin{eqnarray}
		\bm{f}(\bm{k}_3,y) &=& \sum_{\bm{k}_1} \bm{f}(\bm{k}_1,\bm{k}_3-\bm{k}_1,y), \label{eq:fk3conv}\\[0.5em]
		\bm{\tilde{u}}(\bm{k}_3,y) &=& \sum_{\bm{k}_1} \bm{\tilde{u}}(\bm{k}_1,\bm{k}_3-\bm{k}_1,y). \label{eq:uk3conv}
	\end{eqnarray}
\end{subequations}

\begin{figure}
	\centering
	\includegraphics[width=360pt]{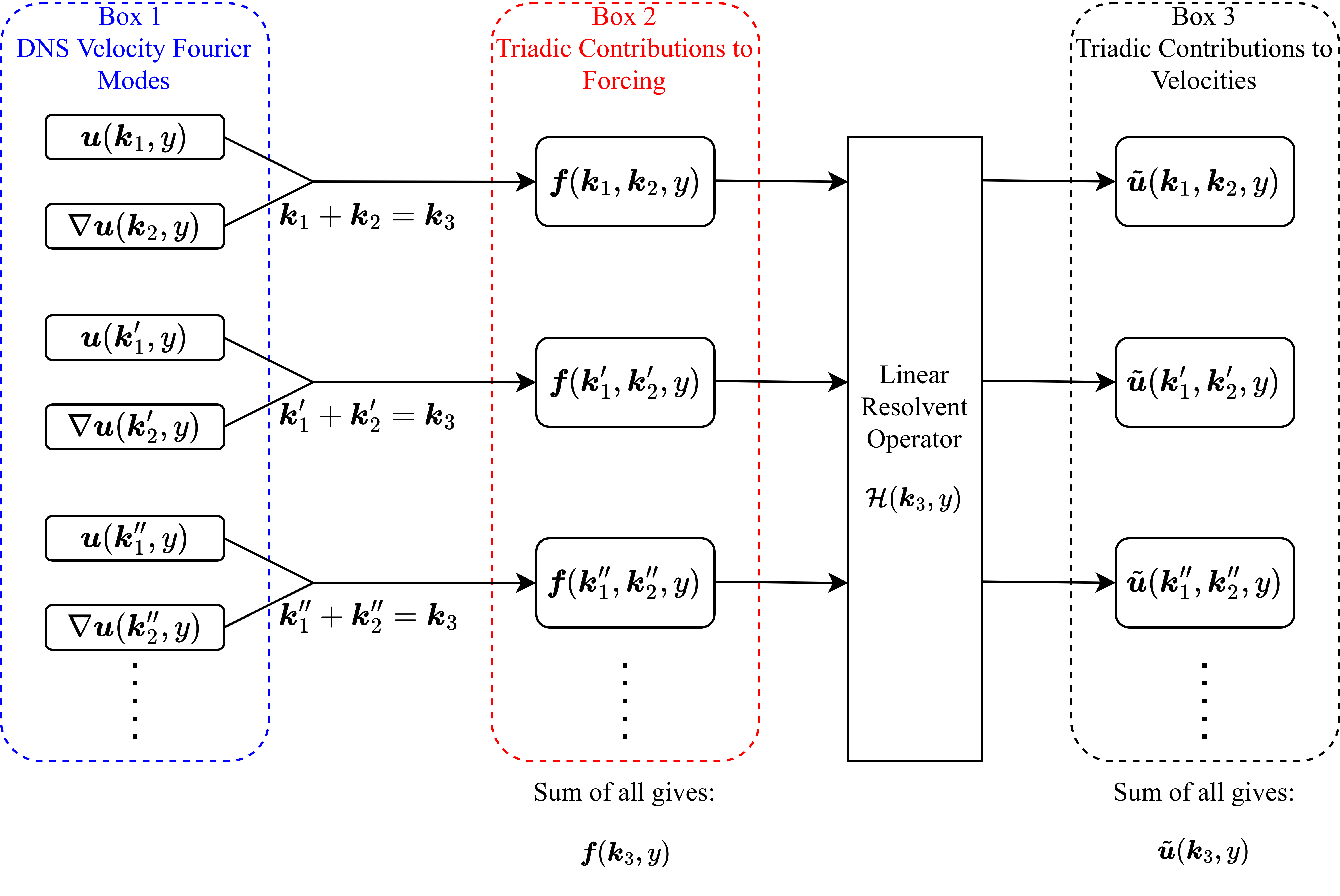}
	\caption{Diagram for the triadic interactions showing individual contributions to the forcing and velocity Fourier modes from interactions between $\bm{k}_1$ and $\bm{k}_2$ that are triadically compatible with $\bm{k}_3$.}
	\label{fig:ti}
\end{figure}

Figure~\ref{fig:ti} gives a visual representation of the relationship between individual scale interactions, the forcing they induce and the associated velocity response. 

\subsection{3-dimensional analyses of forcing and response coefficients}

\begin{figure}
	\centering
	\includegraphics[width=310pt]{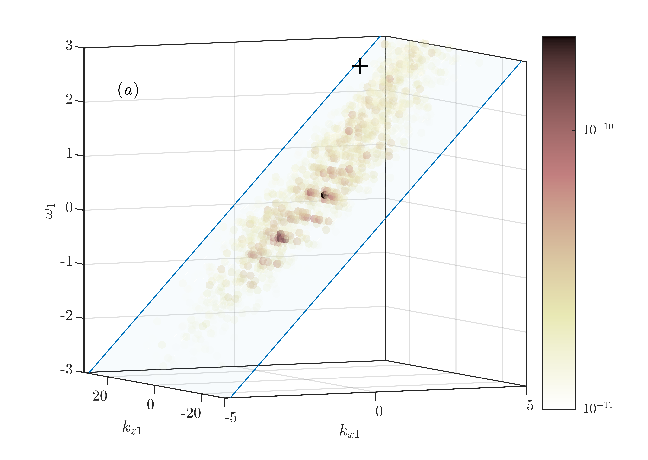}
	\includegraphics[width=310pt]{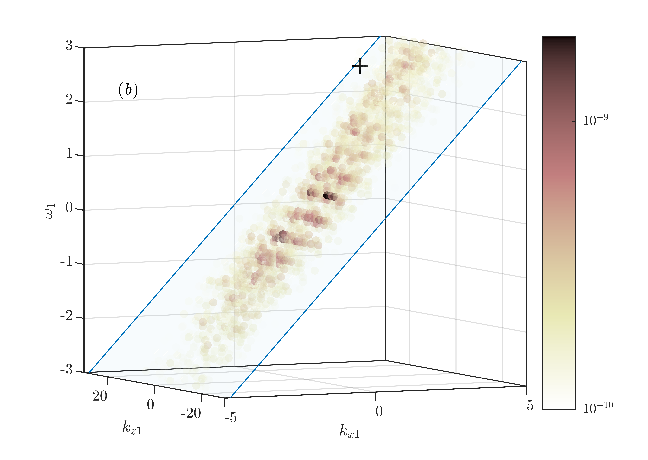}
	\caption{Magnitude of (\textit{a}) $P(\bm{k}_1, \bm{k}_3 - \bm{k}_1)$, the forcing coefficients and (\textit{b}) $R(\bm{k}_1, \bm{k}_3 - \bm{k}_1)$, the response coefficients as functions of $\bm{k}_1  = [k_{x1}, k_{z1}, \omega_1]$ with log scale color bars for $\bm{k}_3 = \bm{k}_1+\bm{k}_2 = [4, \ 28, \ 2.492]$ . The opacity of each marker is also linearly proportional to $\log_{10}$ of the magnitudes. Points with magnitude less than 10\% of the peak values are not plotted. The + markers in both figures denote the location of $\bm{k}_3$, and the blue planes mark the location of $c_1 = \omega_1 / k_{x1} =  c_3 =  0.62$.}
	\label{fig:pr3d}
\end{figure}

For fixed $\bm{k}_3$, the 6-dimensional forcing and response coefficients $P(\bm{k}_1,\bm{k}_2)$ and $R(\bm{k}_1,\bm{k}_2)$ defined in equations~\eqref{eq:defp} and~\eqref{eq:defr} collapse into 3-dimensional quantities, which can be computed and stored with reasonable resources, and can reveal information about the spatio-temporal nature of dominant nonlinear interactions. The magnitude of the forcing and response coefficients are plotted in Figure~\ref{fig:pr3d} as functions of the contributing scale $\bm{k}_1$, noting the (strong) constraint that $\bm{k}_2 = \bm{k}_3 - \bm{k}_1$. 

A first observation is that the majority of the high coefficient values are associated with $k_{x1} \approx 1$, indicating that the large-scale structures play a dominant role in both forcing and response. The largest coefficient magnitudes in the figures are $2.2\e{-10}$ for $P$, and $2.0\e{-9}$ for $R$, reflecting that, while near-wall modes are known to be energetic in a local event, their limited wall-normal reach restricts the integrated energy of such modes. The energy of the selected mode of $\bm{k}_3 = [4, \  28, \ 2.492]$ is $1.3\e{-8}$. As a result, the largest magnitude of $R$, $2.0\e{-9}$ corresponds to 15\% of the total modal energy level, and thus the associated nonlinear interaction is a major contributor towards this representative mode of the near-wall cycle. 

\begin{figure}
	\centering
	\includegraphics[width=380pt]{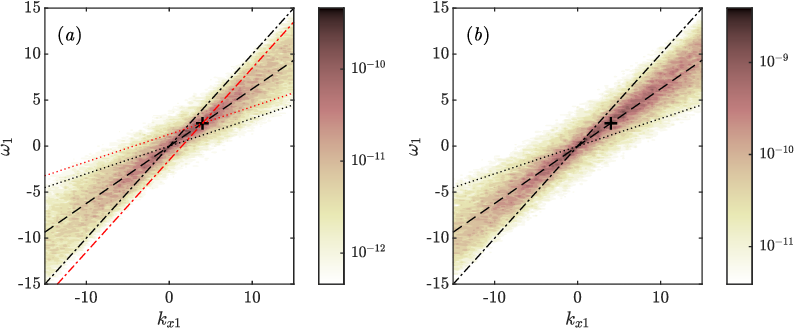}
	\caption{Magnitude of (\textit{a}) $\sum_{k_{z1}} P(\bm{k}_1, \bm{k}_3 - \bm{k}_1)$, the forcing coefficients summed in $k_{z1}$ and plotted as functions of $k_{x1}$ and $\omega_1$ for $\bm{k}_3 = [4, \ 28, \ 2.492]$, and (\textit{b}) $\sum_{k_{z1}} R(\bm{k}_1, \bm{k}_3 - \bm{k}_1)$, the response coefficients. The black dash-dotted lines in both figures mark the wavespeed $c_1 = \omega/k_x = 1$; the black dashed lines for $c_1 = c_2 = c_3 = 0.63$, the wavespeed of the selected $\bm{k}_3$; and the black dotted lines for $c_1 = 0.3$ ($c_1^+ = 6$). The + markers in both figures denote the locations of $\bm{k}_3$. Subplot (\textit{a}) includes an additional set of red lines, with the red dash-dotted line for $c_2 = 1$ and red dotted line for $c_2 = 0.3$ ($c_2^+ = 6$).}
	\label{fig:pr2d}
\end{figure}

Further, it can be observed that most of the high magnitude coefficients (both forcing and response) reside near a single plane with an almost constant $c_1 = \omega_1 / k_{x1}$. This is especially evident in Figure~\ref{fig:pr2d}, which show the same data summed in $k_{z1}$ and plotted as contour plots in the $k_{x1}-\omega_1$ plane. Three different wavespeeds are also shown for reference: $c_1 = \omega_1/k_{x1} = 1$, where the wavespeed matches the center-line velocity; $c_1 = c_3 = 0.63$, the wavespeed of the selected $\bm{k}_3$; and $c_1 = 0.3$ ($c_1^+ = 6$). It can be observed that most of the energetic regions are bounded between $c_1^+ = 6$ and $c_1 = 1$, centering roughly around $c_1 =  c_3 = 0.63$. Note that $c_1 = c_3$ if and only if $c_2 = c_3$ from the triadic compatibility constraint (except for $k_x = 0$ where $c$ is no longer well defined).

\begin{figure}
	\centering
	\includegraphics[width=380pt]{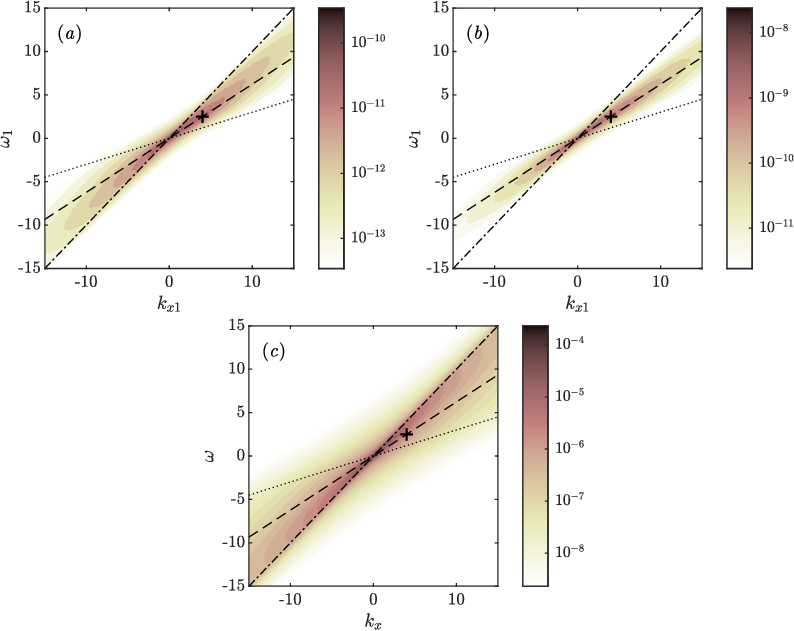}
	\caption{Contour plots of (\textit{a}) $\sum_{k_{z1}} E_f(\bm{k}_1, \bm{k}_3 - \bm{k}_1)$, the forcing energy, and (\textit{b}) $\sum_{k_{z1}} E_u(\bm{k}_1, \bm{k}_3 - \bm{k}_1)$, the response energy, generated by the interactions between $\bm{k}_1$ and $\bm{k}_3 - \bm{k}_1$ summed in $k_{z1}$ and plotted as a function of $k_{x1}$ and $\omega_1$ for $\bm{k}_3 = [4, \ 28, \ 2.492]$. Subplot (\textit{c}) is the contour of $\sum_{k_{z}} E_u(\bm{k})$, the $y$ integrated kinetic energy of the DNS Fourier modes summed in $k_z$ and plotted as a function of $k_{x}$ and $\omega$. The three lines mark the same wavespeeds as in Figure~\ref{fig:pr2d}: dash-dotted lines for $c_1 = 1$, dashed lines for $c_1 = c_2 = c_3 = 0.63$, and dotted lines for $c_1^+ = 6$. }
	\label{fig:efe2d}
\end{figure}

To understand the reason behind the dominance of triadic interactions with  $c_1 \approx c_2 \approx c_3$, we examine the energy (rather than the projections) of the forcing and response generated by the triads $\bm{k}_1$ and $\bm{k}_2 = \bm{k}_3 - \bm{k}_1$. In Figure~\ref{fig:efe2d}(\textit{a}), the energy $E_f(\bm{k}_1, \bm{k}_2)$ and \ref{fig:efe2d}(\textit{b}), the energy $E_u(\bm{k}_1, \bm{k}_2)$ are 
plotted in $k_{x1}-\omega_1$ plane similar to Figure~\ref{fig:pr2d} with the same wavespeeds marked.  $E_u(\bm{k})$, the $y$-integrated turbulent kinetic energy of the DNS velocity Fourier modes, as defined in equation~\eqref{eq:ek}, is shown in Figure~\ref{fig:efe2d}(\textit{c}) to provide context of the energy distribution across all interactions.
Comparing the three figures, it can be observed that the energetic regions of the triadically generated forcing $E_f(\bm{k}_1, \bm{k}_2)$ are located in the vicinity of $c_1 = c_2 = c_3$, while the triadically generated response $E_u(\bm{k}_1, \bm{k}_2)$ lie even closer to $c_1 = c_2 = c_3$.
By contrast, the most energetic regions of full DNS spectrum in Figure~\ref{fig:efe2d}(\textit{c}) are located much closer to $c = 1$, reflecting the energetic dominance of fast-moving modes that have a large wall-normal extent and therefore a large integral contribution to the energy \citep[see, e.g.,][]{Bourguignondecomp13}.

The triadically generated forcing $\bm{f}(\bm{k}_1,\bm{k}_2,y)$ is the product between the velocity modes at $\bm{k}_1$ and the gradient of velocity modes at $\bm{k}_2$. The spatial $y$-localization of the modes around their respective critical layers shown in Figure~\ref{fig:spec} provides one simple insight into the dominant response when modes with $\bm{k}_1$, $\bm{k}_2$ and $\bm{k}_3$ are co-located in $y$.

\subsection{Relative importance of forcing weights and the action of the resolvent}
In this section, we utilize the linear resolvent operator to assess the relative importance of the forcing weights on individual resolvent modes in a triadic interaction and the action of the resolvent in generating the velocity response.


Following Section~\ref{sec:formulation}, the nonlinear weights $\chi_q(\bm{k}_3)$ obtained by projecting the forcing Fourier modes $\bm{f}(\bm{k}_3,y)$ onto the resolvent forcing modes $\phi_q(\bm{k}_3,y)$ are defined as
\begin{equation}
	\chi_q(\bm{k}_3) = \innerproduct{\phi_q(\bm{k}_3,y)}{\bm{f}(\bm{k}_3,y)}.
\end{equation}
Similarly, the contribution to the weight on the $q$-th resolvent forcing mode from individual triadic interactions, $\bm{f}(\bm{k}_1,\bm{k}_2,y)$, can also be defined as
\begin{equation}
	\chi_q(\bm{k}_1,\bm{k}_2) = \innerproduct{\phi_q(\bm{k}_1+\bm{k}_2,y)}{\bm{f}(\bm{k}_1,\bm{k}_2,y)}.
\end{equation}
Thence, forcing and response associated with an individual triadic interaction can be written in terms of weighted resolvent forcing modes as
\begin{subequations}
	\begin{eqnarray}
		\bm{f}(\bm{k}_1,\bm{k}_2,y) &=& \sum_q \chi_q(\bm{k}_1,\bm{k}_2) \phi_q(\bm{k}_1+\bm{k}_2,y), \label{eq:reconstrcut3}\\[0.5em]
		\bm{\tilde{u}}(\bm{k}_1,\bm{k}_2,y) &=& \sum_q   \chi_q(\bm{k}_1,\bm{k}_2) \sigma_q(\bm{k}_1+\bm{k}_2)\psi_q(\bm{k}_1+\bm{k}_2,y). \label{eq:reconstrcut4}
	\end{eqnarray}
\end{subequations}
Finally, the following property can also be obtained from equation~\eqref{eq:fk3conv}:
\begin{equation}
	\chi_q(\bm{k}_3) =  \sum_{\bm{k}_1} \chi_q(\bm{k}_1,\bm{k}_3-\bm{k}_1).
\end{equation} 

Thus, $\chi_q(\bm{k}_3)$ describes the amount of the $q$-th resolvent forcing mode contained in the DNS forcing Fourier mode at $\bm{k}_3$, while $\chi_q(\bm{k}_1,\bm{k}_3-\bm{k}_1)$ is the amount contributed towards the $q$-th resolvent forcing mode by the interaction between $\bm{k}_1$ and $\bm{k}_2 = \bm{k}_3 - \bm{k}_1$. Similarly $\sigma_q(\bm{k}_3)\chi_q(\bm{k}_3)$ describes the amount of the $q$-th resolvent response mode contained in the DNS velocity Fourier mode at $\bm{k}_3$, while $\sigma_q(\bm{k}_3)\chi_q(\bm{k}_1,\bm{k}_3-\bm{k}_1)$ is the amount contributed towards the $q$-th resolvent response mode by the interaction between $\bm{k}_1$ and $\bm{k}_2$. 


Applying these definitions in $P(\bm{k}_1, \bm{k}_2)$ in equation~\eqref{eq:defp} and $R(\bm{k}_1, \bm{k}_2)$ in equation~\eqref{eq:defr} and for the energy in equations~\eqref{eq:ef} and \eqref{eq:eu} yields
\begin{subequations}
	\begin{eqnarray}
		P(\bm{k}_1, \bm{k}_2) &=& \sum_q \expect{\chi_q^*(\bm{k}_1+\bm{k}_2) \chi_q(\bm{k}_1,\bm{k}_2)}, \label{eq:p_resolvent} \\[0.5em]
		R(\bm{k}_1, \bm{k}_2) &=& \sum_q \sigma_q^2(\bm{k}_1+\bm{k}_2)  \expect{\chi_q^*(\bm{k}_1+\bm{k}_2) \chi_q(\bm{k}_1,\bm{k}_2)}, \label{eq:r_resolvent}
	\end{eqnarray}
\end{subequations}
and
\begin{subequations}
	\begin{eqnarray}
		E_f(\bm{k}_1, \bm{k}_2) 
		&=& \sum_q \expect{ \abs{\chi_q(\bm{k}_1,\bm{k}_2)}^2}, \\[0.5em]
		E_u(\bm{k}_1, \bm{k}_2) 
		&=& \sum_q \sigma_q^2(\bm{k}_1+\bm{k}_2) \expect{  \abs{\chi_q(\bm{k}_1,\bm{k}_2)}^2}.
	\end{eqnarray}
\end{subequations}

Note that $\sigma_q$ can be moved outside of the expected value operator as these are quantities computed from the deterministic resolvent operator \citep[c.f.][]{Towne_Schmidt_Colonius_2018}. 
The linear resolvent operator amplifies the different forcing modes differently, and the effect is captured in the response coefficients $R(\bm{k}_1, \bm{k}_2)$ and response energy $E_u(\bm{k}_1, \bm{k}_2)$ through the magnitudes of the $\sigma_q$. 

\begin{figure}
	\centering
	\includegraphics[width=330pt]{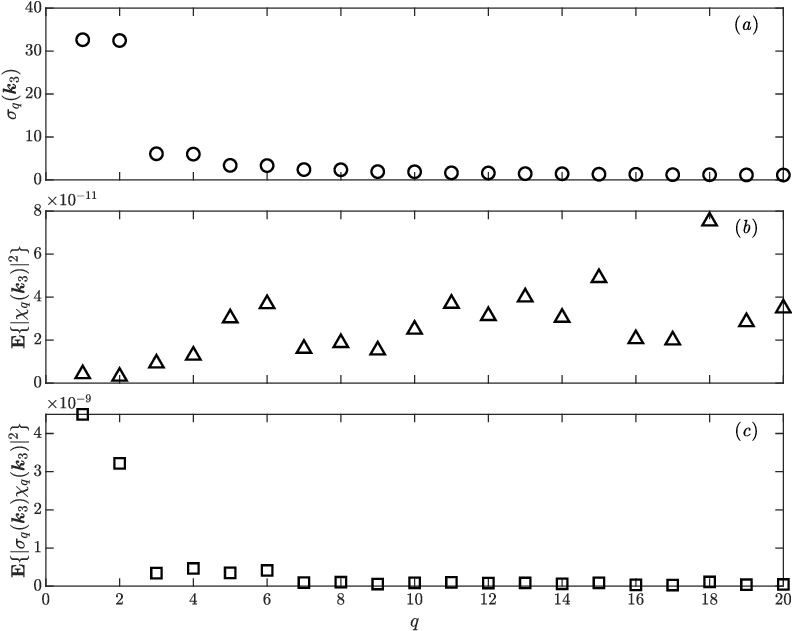}
	\caption{Plots of (\textit{a}) the resolvent singular values $\sigma_q(\bm{k}_3)$, (\textit{b}) the power spectral density of the nonlinear weights $\chi_q(\bm{k}_3)$, obtained by taking the inner product between the resolvent forcing modes and the DNS forcing Fourier mode, and (\textit{c}) the product of the two as a function or the resolvent mode number $q$. }
	\label{fig:k3_sigma_chi}
\end{figure}
For the near-wall cycle mode, $\bm{k}_3 = [4, \  28, \ 2.492]$,
it can be observed from Figure~\ref{fig:k3_sigma_chi}(\textit{a}) that the first two singular values are larger than the rest, showing that the linear resolvent operator predominately amplifies the first two modes and is low-rank at this $\bm{k}_3$. Figure~\ref{fig:k3_sigma_chi}(\textit{b}) shows that the nonlinear weights $\chi_q(\bm{k}_3)$ are roughly the same order of magnitude for all $q$'s shown, with the first few modes having the smallest values and thus contributing only a small portion of the forcing Fourier mode at this $\bm{k}_3$. However, due to the strong amplification of the linear resolvent operator, these first few modes contribute significantly towards the velocity Fourier modes, as evident in Figure~\ref{fig:k3_sigma_chi}(\textit{c}), where it is shown that the first two modes have the largest $\abs{\sigma_q\chi_q}$. The forcing Fourier mode receives contributions from a wide range of resolvent forcing modes, while the velocity Fourier mode is dominated mainly by the first few resolvent response modes due to the low-rank nature of the resolvent. 

These results are consistent with the work of \citet{Morra_Nogueira_Cavalieri_Henningson_2021}, where it is shown that the forcing has significant projection onto the sub-optimal resolvent forcing modes. In fact, the authors showed that the projections onto the first few optimal forcing modes are smaller than the rest, as also observed here. On the other hand, the bulk of the velocity responses is shown to be well approximated using a rank-2 approximation with the first two resolvent modes, which can also be observed in this study.  

\cite{Rosenberg_McKeon_2019} exploited the Helmoholtz decomposition to demonstrate that only the solenoidal component of the forcing effects a velocity response (since the dilatational part can be absorbed into a modified pressure). A large forcing weight, $\chi_q$, can be associated with small solenoidal content of high $q$ forcing modes, leading to a small velocity response as seen here. 

\begin{figure}
	\centering
	\includegraphics[width=300pt]{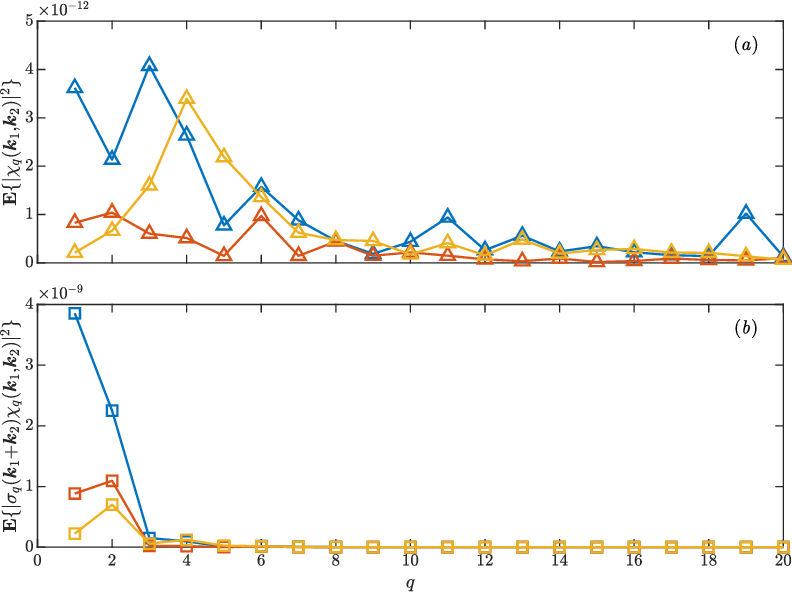}
	\caption{Plots of (\textit{a}) nonlinear weights $\expect{\abs{\chi_q(\bm{k}_1, \bm{k}_2)}^2}$ of triadic interactions between $\bm{k}_1$ and $\bm{k}_2$, and (\textit{b}) the nonlinear weights multiplied by the resolvent singular values $\expect{\abs{\sigma_q(\bm{k}_1+\bm{k}_2)\chi_q(\bm{k}_1, \bm{k}_2)}^2}$ as a function of the resolvent mode number $q$. Blue markers are for triad 1 in table~\ref{tab:triads}, red markers for triad 2, and yellow markers for triad 3.}
	\label{fig:chi_triad}
\end{figure}

Consider the three example pairs of $\bm{k}_1$ and $\bm{k}_2$ that are triadically compatible with $\bm{k}_3 = [4, \ 28, \ 2.492]$ listed in Table~\ref{tab:triads} and the associated forcing and response magnitudes in Figure~\ref{fig:chi_triad}. Triad 1 has strong forcing and response magnitudes. Triad 2 generates a relatively large response from a small forcing, and triad 3 generates a small response even though the forcing magnitude is large.
\begin{table}
	\renewcommand{\arraystretch}{1.2}
	\caption{Three selected triads that contribute to $\bm{k}_3 = \bm{k}_1+\bm{k}_2 = [4, \ 28, \ 2.492]$, with the magnitude of the forcing and response coefficients $\abs{P(\bm{k}_1, \bm{k}_2)}$, $\abs{R(\bm{k}_1, \bm{k}_2)}$ and the energy of the triadically generated forcing and response $E_f(\bm{k}_1, \bm{k}_2)$, $E_u(\bm{k}_1, \bm{k}_2)$.} 
	\label{tab:triads}
	\centering 
	\resizebox{\textwidth}{!}{
	\begin{tabular}{ccccccc} \hline 
			Triad& $\bm{k}_1$ & $\bm{k}_2$ & $\abs{P(\bm{k}_1, \bm{k}_2)}$ & $\abs{R(\bm{k}_1, \bm{k}_2)}$ & $E_f(\bm{k}_1, \bm{k}_2)$ & $E_u(\bm{k}_1, \bm{k}_2)$ \\    \hline 
			1    & $[-0.5, \  2, \ -0.415]$ & $[ 4.5, \ 26, \  2.907]$ & $1.2\e{-10}$ & $1.2\e{- 9}$ & $7.5\e{-11}$ & $6.4\e{-9}$ \\
			2    & $[ 0.5, \ -4, \  0.415]$ & $[ 3.5, \ 32, \  2.077]$ & $1.5\e{-11}$ & $1.9\e{- 9}$ & $2.2\e{-11}$ & $2.0\e{-9}$ \\
			3    & $[-0.5, \  4, \ -0.415]$ & $[ 4.5, \ 24, \  2.907]$ & $1.4\e{-10}$ & $4.3\e{-10}$ & $3.6\e{-11}$ & $1.2\e{-9}$ \\
			\hline
	\end{tabular}}
\end{table}

%
%

For this near-wall mode, the full forcing is shown to have significant projection onto sub-optimal modes, while the full response is relatively low-rank. This is also true on an interaction-by-interaction basis and thus may provide insight into the modeling of Reynolds stress gradients. 

\section{Spatio-temporal characteristics of the nonlinear interactions}\label{sec:results}
We next investigate how individual spatio-temporal scale interactions contribute to the net nonlinear forcing and velocity response across scales and in $y$, examining the variation of the forcing coefficients, $P(\bm{k_1},\bm{k_2})$, and response coefficients, $R(\bm{k_1},\bm{k_2})$, as defined in Equations~\ref{eqs:p} and \ref{eqs:R}. In this section, we focus on the averaged results over all five temporal segments. The analyses were repeated for each temporal segment to study the variation of these coefficients; it was observed (although not included in this paper) that the results described in this section are robust and sufficiently converged for most wavenumbers and frequencies.

\begin{figure} 
	\centering
	\includegraphics[width=382pt]{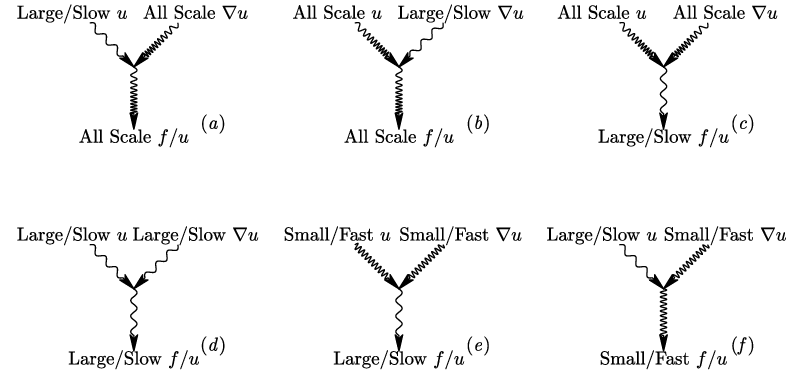}
	\caption{Feynman diagrams depicting the important triadic interactions observed in the forcing or response coefficients. Subplots (\textit{a}), (\textit{b}), (\textit{c}) are for the horizontal, vertical, and diagonal bands of high magnitudes shown in Figures~\ref{fig:pkxrkx}-\ref{fig:pomrom}. Subplots (\textit{d}), (\textit{e}), and (\textit{f}) are for the central region, the top-left corners, and the left or right corners of the horizontal bands. 
    }
	\label{fig:TI}
\end{figure}

Before analyzing the coefficients, it is useful to first define several kinds of possible interactions that will be relevant to the discussion of contributions to the forcing. These are depicted in the Feynman-type diagrams of  Figure~\ref{fig:TI}.

In Figures~\ref{fig:pkxrkx}, \ref{fig:pkzrkz}, and \ref{fig:pomrom}, we will show the magnitudes of the streamwise, spanwise, and temporal (forcing and response) coefficients, respectively, where subplots (\textit{a - d}) show the forcing coefficients, $P_{k_x}$, $P_{k_z}$ and $P_\omega$,  and (\textit{e - h}) the response coefficients, $R_{k_x}$, $R_{k_z}$ and $R_\omega$, for ease of comparison. Different $y$ integration ranges for the inner product defined in equation~\eqref{eq:innerproduct} are shown: integration over all $y$, followed by limits corresponding loosely to the near-wall $(0 \le y^+ \le 30$), overlap ($30\le y^+ \le 200$) and wake ($200 \le y^+ \le 550$) regions.

Each plot has the following structure. The streamwise, spanwise wavenumbers and temporal frequency for the velocity fields, $k_{x1}, k_{z1}, \omega_1$, are on the vertical axis of all figures, and $k_{x2}, k_{z2}, \omega_2$ for the velocity gradients are on the horizontal axis. Lines with a slope of $-1$ correspond to constant $k_{x3} = k_{x1} + k_{x2}$, $k_{z3} = k_{z1} + k_{z2}$, and $\omega_3 = \omega_1 + \omega_2$ for the resulting forcing or response; the extremum values of $k_{x3} = \pm 127.5$, $k_{z3} = \pm 255$, and $\omega_3 = \pm 42.53$ reflect the maximum $k_x$, $k_z$, and $\omega$ retained by the DNS and the temporal Fourier analysis. Four quadrants of $(k_{x1}, k_{x2})$, $(k_{z1}, k_{z2})$, and $(\omega_{1}, \omega_2)$ are shown in subplots (\textit{a, e}) for completeness, while the symmetry discussed in equations~\eqref{eq:hermitiansymmetry} and \eqref{eq:hermitiansymmetry2} is exploited for subsequent subplots (\textit{b - d}) and (\textit{f - h}) in which only $k_{x1},k_{z1},\omega_1 \geq 0$ are shown. To highlight the details of the coefficients, the same logarithmic color scale spanning multiple orders of magnitude is used throughout the different $y$ integration ranges. It should be noted that part of the differences in magnitude between subplots (\textit{b - d}) and between subplots (\textit{f - h}) are simply attributed to the different sizes of the $y$ integration domains. 

The inserts in subplots (\textit{a, e}) of Figures~\ref{fig:pkxrkx}-\ref{fig:pomrom} are representations of the central rectangular regions enclosed in the dotted white lines using linear colorbars. The diagonal white dotted lines with $-1$ slopes in subplots (\textit{a, e}) in Figures~\ref{fig:pkxrkx}-\ref{fig:pomrom} mark the locations of $[k_{x3}, \ k_{z3}, \ \omega_3] = [4, \  28, \ 2.492]$, a representative mode for the near-wall cycle used in section~\ref{sec:traid_nwc}. This mode has a wavespeed of $c_3 = \omega_3/k_{x3} = 0.62$, and corresponds to $\lambda_x^+  = 865$, $\lambda_z^+ = 124$, $\omega_3^+ = 0.0953$, and $c_3^+ = \omega_3^+/k_{x3}^+ = 13$ in inner scales. In the inserts of subplots (\textit{a, e}), the lines are plotted with black dashed lines instead for better contrast in the figures.

Figures~\ref{fig:pkxrkx}-\ref{fig:pomrom} reveal three dominant bands with high magnitudes for all coefficients: the horizontal, vertical, and diagonal bands, centered around $\bm{k}_1, \bm{k}_2$, and $\bm{k}_3\approx0$ respectively. These bands are all nonlinear interactions involving the large/slow scales of the flow and are visually represented by Feynman diagrams in Figures~\ref{fig:TI}(\textit{a - c}). We next analyze these interactions in detail.

\subsection{Triadic interactions in the streamwise direction}
\begin{figure}
	\centering
	\includegraphics[align=c,width=205pt]{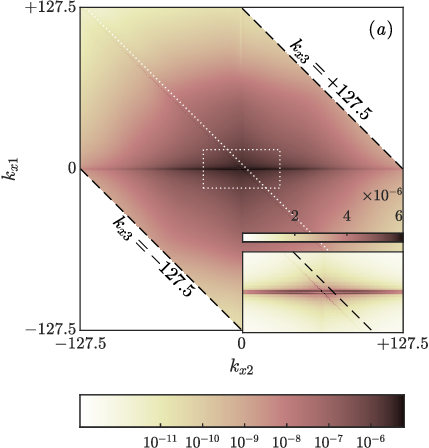}
	\includegraphics[align=c,width=170pt]{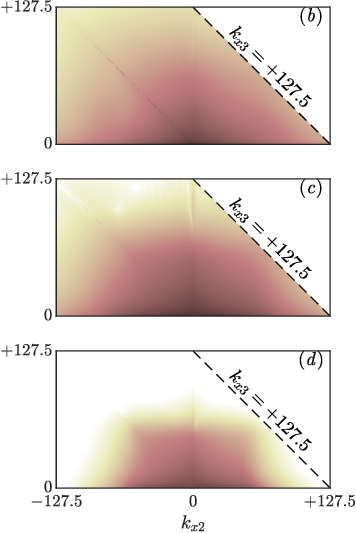}
	\includegraphics[align=c,width=205pt]{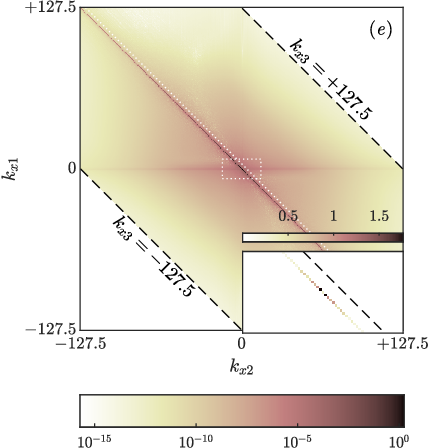}
	\includegraphics[align=c,width=170pt]{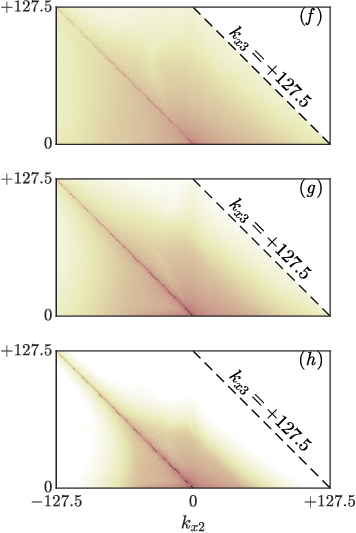}
	\caption{Heatmaps of the magnitude of (\textit{a - d}) the streamwise forcing coefficients $\abs{P_{k_x}(k_{x1}, k_{x2})}$, and (\textit{e - h}) the streamwise response coefficients $\abs{R_{k_x}(k_{x1}, k_{x2})}$. Subplots (\textit{a - d}) except the insert shares the same logarithmic colorbar, same for subplots (\textit{e - h}). The insert in (\textit{a}) and (\textit{e}) corresponds to representations of the rectangular regions enclosed in the dashed white lines using linear scale colorbars. The $y$-integration limits for the inner product in equation~\eqref{eq:innerproduct} are: (\textit{a, e}) all $y^+$; (\textit{b, f}) $y^+ \in (0, 30)$; (\textit{c, g}) $y^+ \in (30, 200)$; and (\textit{d, h}) $y^+ \in (200, 550)$. The streamwise wavenumber for the velocity fields, $k_{x1}$, is on the vertical axis; $k_{x2}$ for the velocity gradient is on the horizontal axis; $k_{x3} = k_{x1} + k_{x2}$ for the resulting forcing and response is constant along lines with slopes of $-1$. The diagonal white dotted lines in (\textit{a, e}) and the black dashed lines in the inserts mark the location of $k_{x3} = 4$ ($\lambda_x^+ = 865$).}
	\label{fig:pkxrkx}
\end{figure}

Consider first the streamwise forcing and response coefficients $\abs{P_{k_x}(k_{x1}, k_{x2})}$ and $\abs{R_{k_x}(k_{x1}, k_{x2})}$ in Figures~\ref{fig:pkxrkx}(\textit{a, e}). A dominant horizontal band is observed for the forcing. This horizontal band, centered around $k_{x1} = 0$, corresponds to the interaction between the streamwise large-scale velocity modes with velocity gradients at all scales, depicted in the Feynman diagram in Figure~\ref{fig:TI}(\textit{a}). By comparison, the same band is present in the response, Figure~\ref{fig:pkxrkx}(e), but with lower amplitude relative to other types of interaction. This is attributable to the action of the resolvent on the forcing from Figure~\ref{fig:pkxrkx}(a).

For the forcing coefficient in Figures~\ref{fig:pkxrkx}(\textit{a - d}), the high values in the horizontal band reflect the concentration of energy in the streamwise large-scale structures at $k_{x1} = 0$ and $\pm0.5$. These modes are also tall, i.e. they have a large wall-normal extent as shown in Figure~\ref{fig:spec}, and therefore are capable of interacting with $k_{x2}$ modes of any size, centered at any $y$ location. The combined effect leads to a significant amount of forcing energy generated across a range of scales by these large-scale modes, manifesting as the energetic horizontal band. Additionally, the values in this horizontal band decay relatively slowly as $\abs{k_{x2}}$ increases. The decreasing energy associated with velocity modes with increasing $\abs{k_{x2}}$ is partially compensated by increasing magnitude of their spatial gradients. 

Finally, comparing the wall-normal variation of the horizontal, $k_{x1}=0$, band across Figures~\ref{fig:pkxrkx}(\textit{b - d}), it can be observed that $P_{k_x}$ exhibits a significant presence across all 3 $y$ ranges due to the tall large-scale modes. The strength of the interactions in the wake region decays, however, for large $\abs{k_{x2}}$, (the interaction type shown in Figure~\ref{fig:TI}(\textit{f})), reflecting the reduction in energy at small scales far from the wall. 

The response coefficients in Figures~\ref{fig:pkxrkx}(\textit{e - h}) show that the horizontal, $k_{x1} = 0$ band does not correspond to the most energetic response. The strong forcing present for $k_{x1} = 0$ at large $k_{x3}$ results in little response energy. This showcases the effect of the linear resolvent operator, which amplifies the small scales less than the large scales and is also consistent with the cascade picture where energy is transferred from the large scales to the small scales through nonlinear interactions and dissipated at the small scales through viscosity \citep{Jimenez_2012}. 

Focusing instead on the different contributors toward the small scales (large $k_{x3}$), it can be observed that the nonlinear interactions between large-scale velocity modes (small $\abs{k_{x1}}$) and small-scale velocity gradients (large $\abs{k_{x2}}$) (Figure~\ref{fig:TI}(\textit{f})) dominate the response. These long-range interactions (in $k_x$) indicate coherence between the large and small scales, consistent with the superposition and modulation mechanisms observed in \citet{marusic2010} and \citet{Andreolli_2023}. 

Lower values of $\abs{P_{k_x}}$ and $\abs{R_{k_x}}$ are observed for $k_{x2}=0$ (vertical band,  representing interactions of the type shown in Figure~\ref{fig:TI}(\textit{b})). The modes contributing to the velocity and velocity gradient are reversed in equations~\eqref{eq:defpkx} and \eqref{eq:defrkx} relative to the $k_{x1}=0$, horizontal band. Spatial derivatives for the large scale are weaker and the energy of the small scales is lower than for the horizontal band, underscoring the asymmetry of interactions within a given triad.

For the diagonal band with $k_{x3} \approx 0$ (interactions of type Figure~\ref{fig:TI}(\textit{c})), a weak signature of 
interactions between two similar size modes generating a streamwise large-scale forcing can be observed for the forcing coefficients in Figures~\ref{fig:pkxrkx}(\textit{a - d}). For the response coefficients in Figures~\ref{fig:pkxrkx}(\textit{e - h}), this diagonal band is stronger, and is the most dominant band of $R_{k_x}$, due to the preferential amplification of the large scale modes by the linear resolvent operator.

Within this diagonal region, forcing coefficient $P_{k_x}$ for interactions between two large scales in the central region (interactions shown in Figure~\ref{fig:TI}(\textit{d})) is stronger than the interactions between two small scales in the corner regions (especially evident in the linear scale inserts). This result is consistent with the findings of \citet{Morra_Nogueira_Cavalieri_Henningson_2021}, where it is demonstrated that large-scale forcing modes are mostly the result of interactions by large-scale structures, although with a relatively weak influence from the interactions between the smaller scales. In addition, the forcing components generated from small-scale interactions are almost non-existent in the outer region (shown in Figure~~\ref{fig:pkxrkx}(\textit{d})), due to the small scales having little energy presence far from the wall. 

On the contrary, the response coefficient $R_{k_x}$, with an energetic diagonal band extending all the way to the small scales in Figures~\ref{fig:pkxrkx}(\textit{e - h}), shows that this type of interaction between small scales affecting the large scales (Figure~\ref{fig:TI}(\textit{e})), has a non-negligible albeit relatively weaker influence on the velocity response. This phenomenon is also demonstrated in the work of~\citet{Illingworth_Monty_Marusic_2018}, where it is shown that including an eddy viscosity into the resolvent framework, which is intended to model the effect of this type of interaction, improves the performance of the resolvent analysis for the large scales. It should be pointed out that Figure~\ref{fig:pkxrkx} examines only the magnitude of the interaction coefficients, where large values show strong importance. The direction of energy transfer, whether energy injection or extraction will be examined later using the phase of the coefficients, and it will be shown that this type of interaction between small scales weakens the spectral TKE of the the large scales, consistent with the energy cascade~\citep{Jimenez_2012}.


Finally, it should be noted that the fact that the three bands have peak values at $k_{x1}, k_{x2}, k_{x3} = 0$ is an artifact of the current numerical simulation. With the current $x$ domain length of $4\pi$, energy from the unresolved large scales manifests as streamwise constant structures at $k_x = 0$, making it the most energetic wavenumber. It should also be pointed out that these are still the mean-subtracted velocity fluctuations, that are streamwise constant, yet non-constant in spanwise direction ($k_z \neq0$) and/or non-constant in time ($\omega \neq 0$). 
As the box size approaches infinity, providing higher and higher resolution for the streamwise wavenumber, these bands are expected to display some width in $k_x$ with multiple wavenumbers very close to $\pm k_{xl}$, the streamwise wavenumber of the most energetic large-scale streak, presumably the very-large-scale motion, displaying high levels of importance.

\subsection{Triadic interactions in the spanwise direction}
\begin{figure}
	\centering
	\includegraphics[align=c,width=195pt]{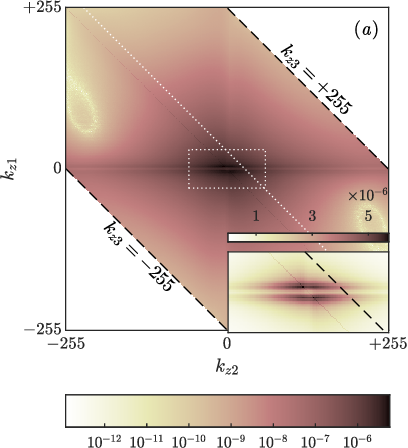}
	\includegraphics[align=c,width=169pt]{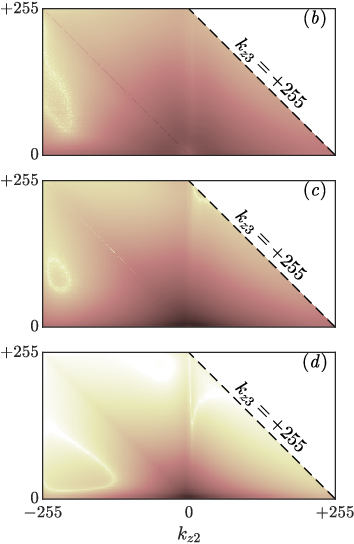}
	\includegraphics[align=c,width=195pt]{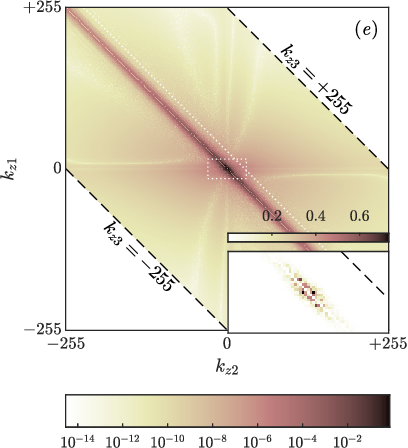}
	\includegraphics[align=c,width=169pt]{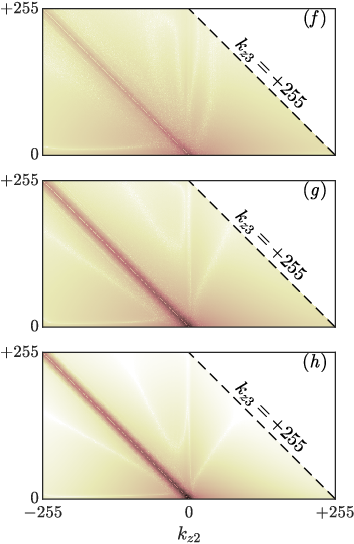}
	\caption{Heatmaps of the magnitude of (\textit{a - d}) the spanwise forcing coefficients $\abs{P_{k_z}(k_{z1}, k_{z2})}$, and (\textit{e - h}) the spanwise response coefficients $\abs{R_{k_z}(k_{z1}, k_{z2})}$ in the same format as Figure~\ref{fig:pkxrkx}. The diagonal white dotted lines in (\textit{a, e}) and the black dashed lines in the inserts mark the location of $k_{z3} = 28$ ($\lambda_z^+ = 124$).}
	\label{fig:pkzrkz}
\end{figure}

The spanwise forcing and response coefficients $\abs{P_{k_z}(k_{z1}, k_{z2})}$ and $\abs{R_{k_z}(k_{z1}, k_{z2})}$ in Figure~\ref{fig:pkzrkz} behave very similarly to the streamwise coefficients, with the exception that the three bands no longer show single-banded structures located around $k_{x}\approx0$, but instead show dual-band structures located at $k_{z} \approx\pm3$. This is as expected since the large-scale structures, which are the most energetic modes in the flow field, with $k_x = 0$ and $k_x = 0.5$ are also most energetic at $k_z \approx \pm 3$. The dual-banded structures are prominent for the energetic horizontal band of $P_{k_z}$ and diagonal band of $R_{k_z}$, while less obvious for the less energetic bands.

The teardrop shaped region at the highest $k_{z2}$ and $k_{z3}$ values observed in the forcing coefficient $P_{k_z}$ in Figures~\ref{fig:pkzrkz}(\textit{a - d}) is likely an artifact of convergence, although note that it is not observed in the response coeffeicient $R_{k_z}$. 
The forcing coefficients in this region have magnitudes that are orders of magnitude smaller than the peak values, making them sensitive to numerical errors. Although the corner areas of the results involving extreme $k_z$ values are less robust quantitatively, the overall structure of the results is expected to remain consistent.


\subsection{Triadic interactions of the temporal frequencies}

\begin{figure}
	\centering
	\includegraphics[align=c,width=202pt]{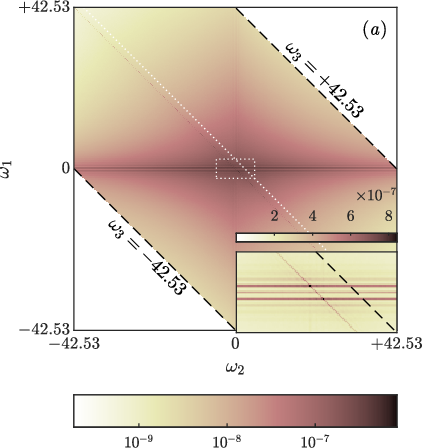}
	\includegraphics[align=c,width=170pt]{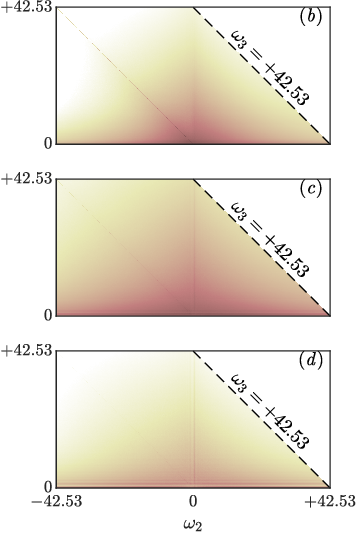}
	\includegraphics[align=c,width=202pt]{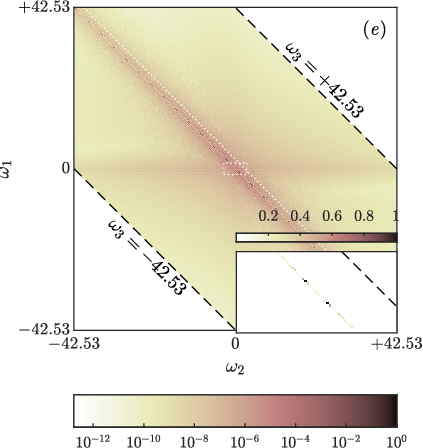}
	\includegraphics[align=c,width=170pt]{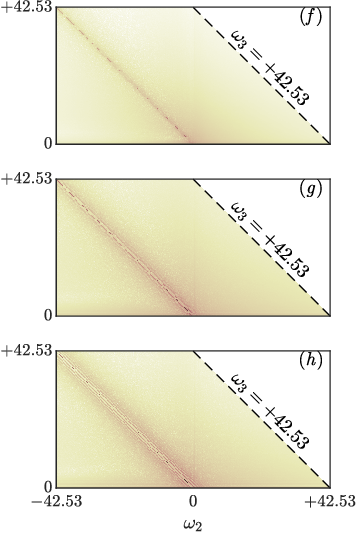}
	\caption{Heatmaps of the magnitude of (\textit{a - d}) the temporal forcing coefficients $\abs{P_{\omega}(\omega_{1}, \omega_{2})}$, and (\textit{e - h}) the temporal response coefficients $\abs{R_{\omega}(\omega_{1}, \omega_{2})}$ in the same format as Figure~\ref{fig:pkxrkx}. The diagonal white dotted lines in (\textit{a, e}) and the black dashed lines in the inserts mark the location of $\omega_{3} = 2.492$ ($\omega_3^+ = 0.0953$).}
	\label{fig:pomrom}
\end{figure}

The overall structure of the frequency forcing and response coefficients $\abs{P_{\omega}(\omega_{1}, \omega_{2})}$ and $\abs{R_{\omega}(\omega_{1}, \omega_{2})}$ plotted in Figure~\ref{fig:pomrom} is again similar to that of the streamwise coefficients, with the exception that the single-banded structures in the streamwise direction now becomes multi-banded. These prominent discrete high-value lines located in the horizontal, vertical, and diagonal bands with $\omega_1, \omega_2, \omega_3 \approx0$ are due to the discreteness in the streamwise wavenumber $k_x$, an artifact of the finite simulation domain length, as similarly demonstrated in \citet{Gomez2014}. The frequency can be related to the phase speed for a given $\bm{k}$ via $\omega = c\cdot k_x$; at a given wavespeed, $c$, an increase in $k_x$ to the next discrete wavenumber, with an increment of $0.5$ fixed by the simulation domain length, will cause $\omega$ to increase by $0.5c$. Therefore, increments in $\omega$ are largest for large scales with high wavespeeds, appearing as discrete lines, and smaller for small scales with low wavespeeds, reflected as the smooth varying background. 

A more detailed analysis reveals that the 5 most prominent lines observed in Figure~\ref{fig:pomrom} for both $P_{\omega}$ and $R_{\omega}$ are located at $\omega\approx0$, $\pm 0.4$, and $\pm0.8$. These lines correspond well with the energetic modes at $k_x=0$, $k_x=\pm 0.5$ (with wavespeeds of $c = \omega/k_x \approx0.8$ shown in Figure~\ref{fig:spec}(\textit{a})) and $k_x= \pm 1$ (with $c \approx0.8$ shown in Figure~\ref{fig:spec}(\textit{b})) respectively. Furthermore, by observing the relative intensities between the smooth background and discrete lines across different $y$ locations, it can be seen that the former is more prominent near the wall in Figures~\ref{fig:pomrom}(\textit{b, f}) where the discreteness is barely visible, while away from the wall in Figures~\ref{fig:pomrom}(\textit{d, h}) the opposite is true. Combined with the tall $y$ extent for large scales and the concentration of energy near the wall for small scales, as demonstrated in Figure~\ref{fig:spec}, these plots confirm that the smooth background mostly shows the triadic interactions between the small scales, while the discrete lines are mostly the result of interactions involving the large scales. Although the discreteness is an artifact of the finite simulation domain length, interactions involving large scales are expected to be important regardless of the domain length.

\subsection{Constructive and destructive triadic interactions}
\begin{figure} 
	\centering
	\includegraphics[width=379pt]{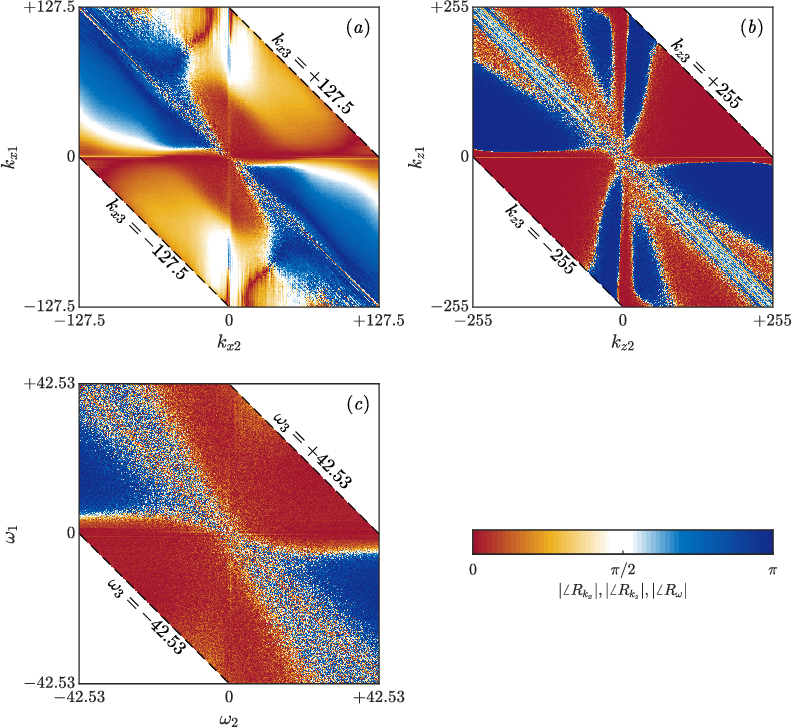}
	\caption{Heatmaps of the absolute values of the phase angles of the response coefficients: (\textit{a}) $\abs{ \angle R_{k_x}(k_{x1}, k_{x2})}$, (\textit{b}) $\abs{ \angle R_{k_z}(k_{z1}, k_{z2})}$, and (\textit{c}) $\abs{ \angle R_{\omega}(\omega_{1}, \omega_{2})}$. Phase angles close to 0 (red) indicate constructive interference, while phase angles close to $\pi$ (blue) indicate destructive interference.}
	\label{fig:rphase}
\end{figure}

Both the forcing and response coefficients are complex numbers, with the magnitude providing information about the importance of a triadic interaction, while the phase provides information about constructive and destructive interference. From equations~\eqref{eq:rkx_sum}-\eqref{eq:rom_sum}, it can be observed that the response coefficients sum to yield the spectral turbulent kinetic energy of the response modes, which are real positive quantities. Therefore, the imaginary parts of the coefficients cancel upon summation, while the real parts provide information about the constructive and destructive contributions to the spectral TKE. A positive real part of $R(\bm{k}_1,\bm{k}_2)$ indicates that the interaction between $\bm{k}_1$ and $\bm{k}_2$ causes an increase in spectral TKE of the response mode at $\bm{k}_3$, while a negative real part indicates a decrease of spectral TKE. To analyze the constructive and destructive interference, we utilized the phase angle of the response coefficients: a phase angle within $(+\pi/2, -\pi/2)$ or the right half plane of the complex plane indicates a positive real part with both positive or negative imaginary part; while a phase angle within $(+\pi/2,+\pi) \cup (-\pi,-\pi/2)$ or the left half plane indicates a negative real part. Since the sign of the imaginary part is of less interest, we utilize the absolute value of the phase angles, where $\abs{\angle R} \in [0,\pi/2)$ indicates constructive interference or in other words, increase of spectral TKE, and $\abs{\angle R} \in (\pi/2,\pi]$ indicates destructive interference or in other words, decrease of spectral TKE. In Figures~\ref{fig:rphase}, the absolute value of the phase of $R_{k_x}$, $R_{k_z}$, and $R_{\omega}$ are plotted, with phase angles close to 0 in red indicating constructive interference and phase angles close to $\pi$ in blue indicating destructive interference.

It should be noted that the constructive and destructive interference studied here are different from typical studies of energy transfer focusing on the turbulent transport term. As noted in equation~\eqref{eq:stke}, a nonlinear energy transfer into a mode at $\bm{k}_3$ by turbulent transport does not necessarily result in an increase in spectral TKE due to the presence of other linear mechanisms. However, in this study, with both the nonlinear energy transfer and linear energy amplification mechanisms studied together, constructive or destructive interference directly indicates an increase or decrease of the spectral TKE, providing a different perspective compared to other energy transfer studies.

The overall structure of the phase plots can be seen as a red hour-glass structure spanning from the bottom-left to the top-right, which is obvious across all three subplots, and a blue hour-glass structure spanning from the top-left to bottom right, which is obvious for $R_{k_x}$ and $R_{\omega}$, but exhibits more complex behavior for $R_{k_z}$. To analyze this structure, we will focus on lines of constant $\bm{k}_3$, which are lines with slopes of $-1$, parallel to the dashed lines marking the extreme values of $\bm{k}_3$. Equation~\eqref{eq:rkx_sum} indicates that summing along this line of constant $k_{x3}$ gives the energy of all modes at $k_{x3}$. Within this line, the central region with small $k_{x1}$ and $k_{x2}$, representing the interactions between the large scales, generally contributes positively to the energy, while the corner regions, representing the interactions between small scales, generally reduce energy. This is consistent with the energy cascade, where turbulent kinetic energy is being generated at the large scales, transferred to small scales through triadic interactions, and dissipated at the small scales by viscosity. For the phase of $R_{\omega}$, the general structure behaves similarly. However, the central region is more fuzzy while generally remaining red, contributing positively towards the energy. This is likely due to the fact that, unlike $k_x$ where only large scales contribute towards low $k_x$, for $\omega$, both large scales with high wavespeeds and small scale with low wavespeeds can contribute to low $\omega$. This phenomenon results in the coexistence of the high-value discrete lines and the smooth varying background in the magnitude of $R_{\omega}$ discussed in the previous section, and mostly likely contributes to the fuzziness in the central region. For the phase of $R_{k_z}$, the general structure remains similar. It is observed that the overall structure of these results is robust and sufficiently converged; however, the details of the transitional regions between constructive and destructive interference require more data to reduce the variations for future studies focusing on these regions.

\section{Quasi-linear and generalized quasi-linear contributions to the forcing and response}
In Figure~\ref{fig:pkxrkx}, we observed three regions of dominant contributions to the forcing and response, all corresponding to triadic interactions involving the streamwise large scales, consistent with the assumptions underlying QL and GQL reductions of the NSE. It should be emphasized that $P_{k_x}$ and $R_{k_x}$ are measurements of the triadic contributions to the total forcing and response using data from the DNS, which is a different dynamical system compared to QL/GQL. Nevertheless, as all are mathematical approximations to the same physical system, the following analyses provide insights into the types nonlinear interactions retained or lost in QL and GQL.

We start by decomposing the velocity $\bm{u}$ and nonlinear forcing $\bm{f}$ in a manner reflecting the QL and GQL restrictions~\citep[e.g.][]{Marston_Chini_Tobias_2016}:
\newcommand*\centermathcell[1]{\omit\hfil$\displaystyle#1$\hfil\ignorespaces}
\begin{alignat}{7}
	\bm{u} &=&& \centermathcell{\bm{\bar{u}}}&&+&&\centermathcell{\bm{\tilde{u}}}&&+&&\centermathcell{\bm{u'}},\\
	\bm{f} &=&& \underbrace{\bm{\bar{f}}}_{k_x = 0}&&+&&\underbrace{\bm{\tilde{f}}}_{0<\abs{k_x}\leq \Lambda} &&+&& \underbrace{\bm{f'}}_{\abs{k_x}>\Lambda}.
\end{alignat}
Here $\bm{\bar{u}}, \bm{\bar{f}}$ are the streamwise averages, i.e. all modes with $k_x=0$. 
$\bm{\tilde{u}}$ and $\bm{\tilde{f}}$ contain the large scales with $k_x$ less than or equal to the cut-off $\Lambda$, i.e. $0<\abs{k_x}\leq \Lambda$, and $\bm{u'},\bm{f'}$ contain the residual, i.e. small scales with $\abs{k_x}>\Lambda$. Note that the spatio-temporal mean profile $\overline{U}$ with $\bm{k}=(k_x, k_z, \omega) = (0, 0, 0)$ appears as part of $\bm{\bar{u}}$ under this decomposition. 

The three terms may be grouped to reflect QL or GQL system formulations. In QL, $\Lambda= 0$, such that $\bm{\bar{u}}$ represents the streamwise constant base flow and $\bm{u'}$ the perturbation, while $\bm{\tilde{u}}$ is zero. For GQL, the large-scale base flow consists of all contributions with $k_x \le \Lambda$, i.e. $\bm{\bar{u}}+\bm{\tilde{u}}$, and $\bm{u'}$ is the perturbation.

\begin{figure} 
	\centering
	\includegraphics[width=233pt]{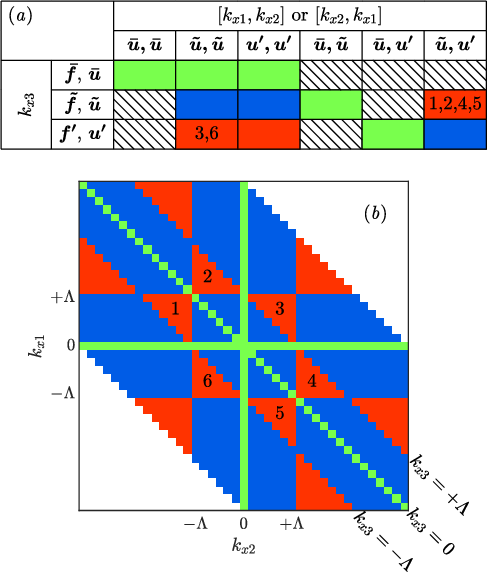}
	\caption{The regions of triadic interactions included in QL/GQL in (\textit{a}) tabular form and (\textit{b}) graphical form for comparison with Figure~\ref{fig:pkxrkx}. The color for the table cells and figure are green for triadic interactions resolved in both QL and GQL; blue for additional triadic interactions included in GQL but not in QL; and red for triadic interactions modeled or neglected in both QL and GQL. Hashed cells in the table indicate prohibited interactions. Six special red triangular regions in the center are labeled 1-6, marking the regions that expand as $\Lambda$ increases. The interaction type of each triangular region is also marked in the corresponding cell in the table.}
	\label{fig:GQLRegion} 
\end{figure}

Figure~\ref{fig:GQLRegion}(\textit{a}) shows the triadic interactions permitted by QL/GQL in a tabular form, with the six possibilities for the velocity or the velocity gradient listed in the six columns, and the resulting forcing or response listed in the three rows. These regions of interactions are also plotted in Figure~\ref{fig:GQLRegion}(\textit{b}) in a $k_{x1}$ vs $k_{x2}$ plane similar to Figure~\ref{fig:pkxrkx}. In this figure, the color green indicates interactions resolved in both QL/GQL and corresponds to three lines with $k_{x1}, k_{x2}$, or $k_{x3}=0$ in Figure~\ref{fig:GQLRegion}(\textit{b}). The color blue indicates additional interactions included in GQL but not in QL, and in the limiting cases for GQL with $\Lambda = 0$, for which GQL is equivalent to QL, the blue regions disappear, and the triple decomposition collapses to a double decomposition. The color red indicates interactions that are modeled or neglected in both QL/GQL. In the limiting case for  GQL with $\Lambda\geq \max(k_x) = 127.5$, the red regions disappear, as all the nonlinear interactions included in the DNS are also included in GQL. Finally, the hashed cells in Figure~\ref{fig:GQLRegion}(\textit{a}) indicate non-resonant, prohibited interactions; for example, the interaction of $\bm{\bar{u}}$ and $\bm{\bar{u}}$ (both $k_x = 0$) can contribute to $\bm{\bar{f}}, \bm{\bar{u}}$ ($k_x = 0$), but not $\bm{\tilde{f}}, \bm{\tilde{u}}$ nor $\bm{f'}, \bm{u'}$. From Figure~\ref{fig:GQLRegion}(\textit{a}), it can be observed that all triadic interactions contributing to $\bm{\bar{f}}, \bm{\bar{u}}$ are resolved in QL/GQL, while for $\bm{\tilde{f}}, \bm{\tilde{u}}$ and $\bm{{f'}}, \bm{{u'}}$ only part of the triadic interactions are resolved.

\begin{figure} 
	\centering
	\includegraphics[width=0.30\linewidth]{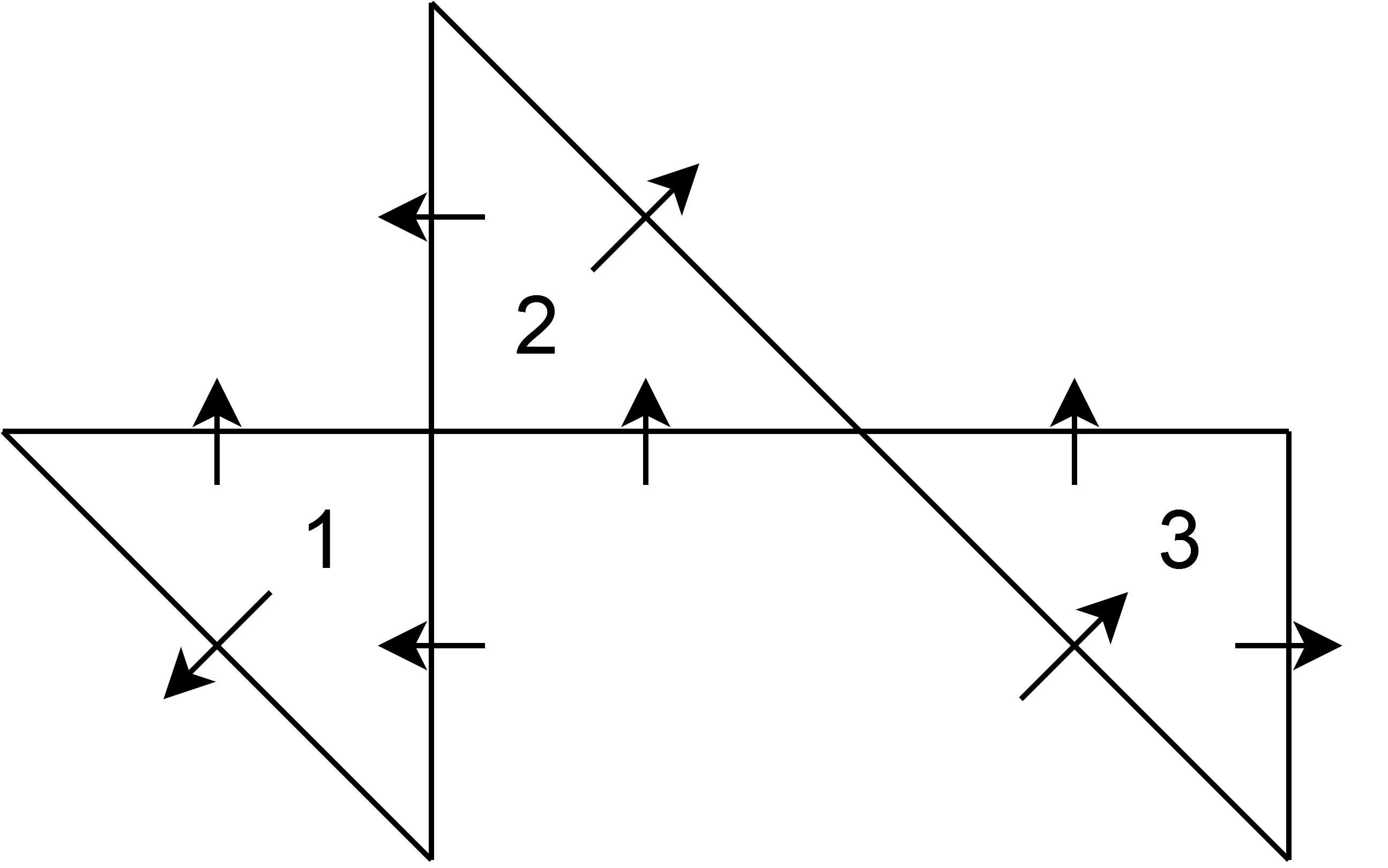}
	\caption{The direction of boundary movement for the red triangular regions 1-3 in the Figure~\ref{fig:GQLRegion} as $\Lambda$ increases. Triangular regions 4-6 are the mirror images of 1-3 and are omitted in this sketch.}
	\label{fig:GQLTriangles} 
\end{figure}

Upon close inspection of the six red triangular regions in the center (labeled 1-6 in Figure~\ref{fig:GQLRegion}(\textit{b})), it can be observed that the three boundaries of these triangles do not all move inwards as $\Lambda$ increases. As $\Lambda$ changes, an inward-moving boundary turns red regions into blue, indicating the inclusion of more triadic interactions in GQL. On the other hand, an outward-moving boundary turns a previously blue region red, indicating a loss of some previously included triadic interactions. The direction of movement of the boundary of triangular regions 1-3 are sketched in Figure~\ref{fig:GQLTriangles}. As $\Lambda$ increases, only one boundary of each triangular region 1-6 moves inwards, and these triangular regions, having side lengths of $\Lambda$, increase in size and move further from the center. These triangular regions eventually reach the boundary of the figure (maximum $k_x$ retained by the DNS), then start to decrease in size as portions of them are now outside of the figure. When $\Lambda$ reaches $\max(k_x) = 127.5$, they move completely out of the figure and the GQL becomes equivalent to the DNS. The important consequence of this is that as $\Lambda$ increases, although more triadic interactions are being included globally, a portion of the previously included ones are now lost. Depending on the relative importance of the newly included and lost interactions, the increase in $\Lambda$ could cause non-monotonic performance changes under certain conditions.

Comparing Figure~\ref{fig:GQLRegion}(\textit{b}) with Figure~\ref{fig:pkxrkx}, it can be seen that the QL assumptions do indeed restrict resolved interactions to those corresponding to large forcing and response coefficients in the DNS. The blue regions in Figure~\ref{fig:GQLRegion}(\textit{b}), which represent the additional interactions resolved in GQL, also correspond to large contributions to the overall forcing and response in Figure~\ref{fig:pkxrkx}. The fractional contribution of GQL-permitted interactions to the total DNS forcing and response for varying $\Lambda$ can be quantified with the following ratios:
\begin{eqnarray}
	\rho_{f}(\Lambda) &=& \frac{\sum_{GQL(\Lambda)}  P_{k_x}(k_{x1}, k_{x2})}{\sum_{GQL(\infty)}  P_{k_x}(k_{x1}, k_{x2})},\\[0.5em]
	\rho_{r}(\Lambda) &=& \frac{\sum_{GQL(\Lambda)}  R_{k_x}(k_{x1}, k_{x2})}{\sum_{GQL(\infty)}  R_{k_x}(k_{x1}, k_{x2})},
\end{eqnarray}
where $\sum_{GQL(\Lambda)}$ indicates a summation in the $k_{x1}, k_{x2}$ regions resolved by GQL with the parameter $\Lambda$ (a summation over the green and blue regions in Figure~\ref{fig:GQLRegion}(\textit{b})). As  $\Lambda \to \infty$, the GQL assumptions admit the equivalent range of interactions to the DNS, with $\rho_{f}(\Lambda \to \infty) = \rho_{r}(\Lambda \to \infty) = 1$, and $\Lambda = 0$ indicates no blue region, with GQL equivalent to QL, and the ratios describing the fractional energy captured by QL. 

\begin{figure} 
	\centering
	\includegraphics[width=275pt]{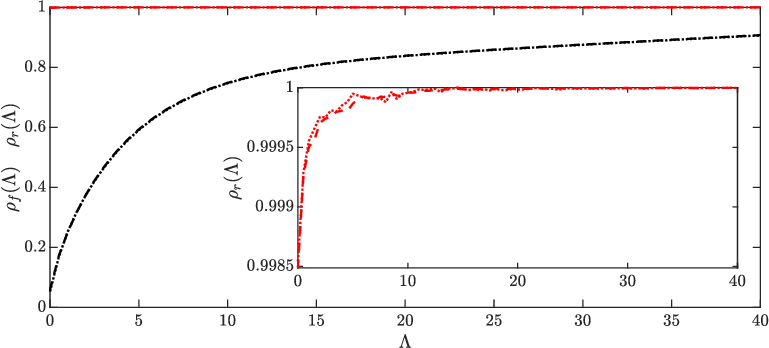}
    \caption{Fraction of total DNS forcing and response energy captured by interactions obeying GQL assumptions for various values of $\Lambda$. The black lines are $\rho_{f}(\Lambda)$ for the forcing and the red lines are $\rho_{r}(\Lambda)$ for the response. For both the forcing and response, the dashed lines are results computed using the coefficients averaged over all five temporal segments, while the dotted lines are the results computed using only the first temporal segment. The insert is a zoomed in view of $\rho_r(\Lambda)$.}
	\label{fig:GQLRatio}
\end{figure}

The ratios $\rho_{f}(\Lambda)$ and $\rho_{r}(\Lambda)$ are plotted in Figure~\ref{fig:GQLRatio}. It can be seen that for QL ($\Lambda = 0$), a small amount of forcing energy is captured while almost all the response energy is already captured. This is due to the fact that almost all the response energy is concentrated at $k_{x3} = 0$, as shown in the insert of Figure~\ref{fig:pkxrkx}(\textit{e}), and thus QL is expected to perform well for this flow. However, as discussed in the previous section, the concentration at $k_{x3} = 0$ is likely the result of a small simulation domain size. With larger domains properly resolving the streamwise large scales, the concentration is expected to be located at small but non-zero $k_{x3}$, which necessitates the use of GQL.	In addition, $\rho_{f}(\Lambda)$ and $\rho_{r}(\Lambda)$ converge rapidly, due to the summation over regions of $P_{k_x}$ and $R_{k_x}$, despite requiring a large number of snapshots for the convergence of $P_{k_x}$ and $R_{k_x}$ themselves. Small differences are observed in Figure~\ref{fig:GQLRatio} when $\rho_{f}$ and $\rho_{r}$ are computed using the first temporal segment rather than the average of all five temporal segments, and therefore $\rho_{f}$ and $\rho_{r}$ can be approximated with a short statistically steady DNS run.

\begin{figure} 
	\centering
	\includegraphics[width=275pt]{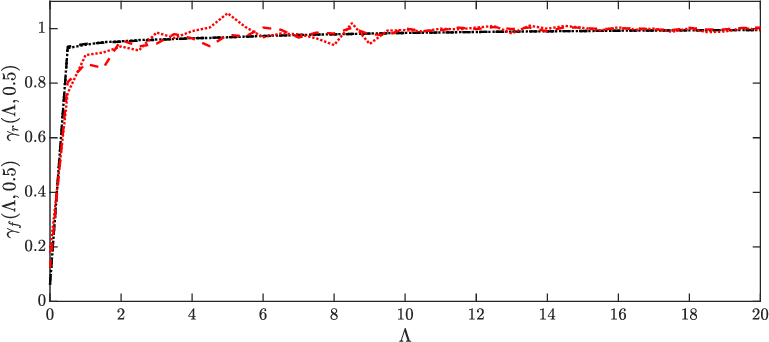}
	\includegraphics[width=275pt]{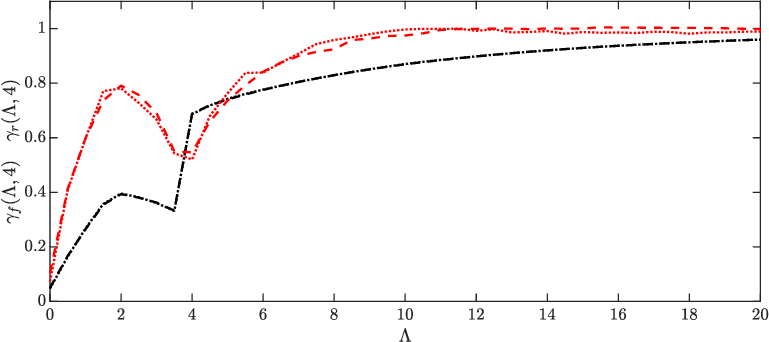}
    \caption{Fraction of total DNS forcing and response energy captured by interactions obeying GQL assumptions for various values of $\Lambda$ and restricted to $k_{x3} = k_{x1} + k_{x2} = 0.5$, a representative large scale (top) and $k_{x3} = k_{x1} + k_{x2} = 4$ ($\lambda_x^+ \approx 900$), the peak of the near-wall cycle (bottom). The black lines are $\gamma_{f}(\Lambda, 0.5)$ for the forcing and the red lines are $\gamma_{r}(\Lambda,0.5)$ for the response. For both the forcing and response, the dashed lines are results computed using the coefficients averaged over all 5 temporal segments, while the dotted lines are the results computed using only the first temporal segment.}
	\label{fig:GQLRatio2}
\end{figure}

With the previous analysis of GQL regions for all $k_{x3}$ dominated by the mode at $k_{x3} = 0$, we now perform the analysis again for specific values of $k_{x3}$. The two energy ratios are redefined:
\begin{eqnarray}
	\gamma_{f}(\Lambda,k_{x3}) &=& \frac{\sum_{GQL(\Lambda), k_{x1} + k_{x2} = \pm k_{x3}}  P_{k_x}(k_{x1}, k_{x2})}{\sum_{GQL(\infty), k_{x1} + k_{x2} = \pm k_{x3}}  P_{k_x}(k_{x1}, k_{x2})},\\[0.5em]
	\gamma_{r}(\Lambda,k_{x3}) &=& \frac{\sum_{GQL(\Lambda), k_{x1} + k_{x2} = \pm k_{x3}}  R_{k_x}(k_{x1}, k_{x2})}{\sum_{GQL(\infty), k_{x1} + k_{x2} = \pm k_{x3}}  R_{k_x}(k_{x1}, k_{x2})}.
\end{eqnarray}
The summation $\sum_{GQL(\Lambda), k_{x1} + k_{x2} = \pm k_{x3}}$, is still the summation in the $k_{x1}, k_{x2}$ regions resolved by GQL with the parameter $\Lambda$ (green and blue regions), with the added restriction of $k_{x1} + k_{x2} = \pm k_{x3}$ (along two lines with slopes of -1 corresponding to constant $\pm k_{x3}$). The resulting ratios are plotted in Figure~\ref{fig:GQLRatio2} for $k_{x3} = 0.5$, a representative large scale and $k_{x3} = 4$ ($\lambda_x^+ \approx 900$), the peak of the near wall cycle.

For $k_{x3} = 0.5$ in Figure~\ref{fig:GQLRatio2}, it can be observed that at $\Lambda = 0$ (QL), very little energy for both the forcing and response are captured. This is expected as $k_x = 0.5$ is not contained in the large scale base flow for $\Lambda = 0$, and the only triadic interactions included are $(k_{x1},k_{x2}) = (0,0.5)$ and $(0.5,0)$, which constitutes a small fraction of energy for $k_{x3} = 0.5$. Both $\gamma_{f}(\Lambda,0.5)$ and $\gamma_{r}(\Lambda,0.5)$ immediately jump above 0.8 starting from $\Lambda = 0.5$. For $\Lambda \geq 0.5$, the $k_x = 0.5$ modes are included in the large-scale base flow, and almost all triadic interactions contributing to $k_{x3} = 0.5$ are included except for the pairs $(k_{x1},k_{x2}) = (\Lambda+0.5,-\Lambda)$ and $(-\Lambda,\Lambda+0.5)$ (the tips of triangles 2 and 4 in Figure~\ref{fig:GQLRegion}(\textit{b})). As $\Lambda$ increases, this neglected pair of interactions moves towards less energetic regions and the energy ratios quickly converge to 1, with some overshoots due to the omission of destructive interferences.  

For the restriction to interactions associated with $k_{x3} = 4$, corresponding to the peak of the near wall cycle, more interesting behavior is observed. The ratios start off at a low value for $\Lambda = 0$ for the same reason as the previous case, then steadily increase until $\Lambda = 2$ before decreasing until $\Lambda = 4$, and finally increasing steadily until convergence toward $\gamma_f, \gamma_r = 1$. We first compute the range of triadic interactions \textit{not} included for given $\Lambda$:
\begin{equation}
	k_{x1} \in \left\{ 
	\begin{array}{ll}
		( -\infty,-\Lambda) \cup (\Lambda, 4-\Lambda) \cup (4+\Lambda, \infty)&\text{for } \Lambda \in (0,2)\\
		( -\infty,-\Lambda) \cup [4-\Lambda, \Lambda] \cup (4+\Lambda, \infty)&\text{for } \Lambda \in [2, 4)\\
		(\Lambda, \Lambda+4] \cup [-\Lambda, 4-\Lambda)                       &\text{for } \Lambda \in [4, \infty)\\
	\end{array}
	\right. .
\end{equation}
It can be seen that for $\Lambda \in (0,2)$, all three ranges shrink in size as $\Lambda$ increases, indicating more triadic interactions are being steadily added while none are lost. In addition, due to the small $\Lambda$, the included regions almost exclusively contribute to constructive interference, resulting in monotonically increasing $\gamma_f$ and $\gamma_r$. For $\Lambda \in [2, 4)$ however, the triangle 3 in Figure~\ref{fig:GQLRegion}(\textit{b}) is now one of the regions not included for $k_{x3} = 4$. As $\Lambda$ increases, this red region increases in size, losing triadic interactions and causing $\gamma_f$ and $\gamma_r$ to decrease. Finally, for $\Lambda\geq 4$, $k_{x3} =4$ is now included in the base flow as one of the large scales, although the ranges of not included $k_{x1}$ remain constant in size (going across triangles 2 and 4 in Figure~\ref{fig:GQLRegion}(\textit{b})), they get pushed out to less energetic regions, resulting in increasing $\gamma_f$ and $\gamma_r$, eventually converging to 1. This phenomenon is, in fact, the standard behavior for most (if not all) $k_{x3} \geq 1.5$, where the ratios increase for $\Lambda \in (0,k_{x3}/2)$, decrease for $\Lambda \in [k_{x3}/2, k_{x3})$ and increase again for $\Lambda \in [k_{x3}, \infty)$. $k_{x3} = 0.5$ and $1$ do not behave like this due to the non-existence of the first two ranges of $\Lambda$. However, it should be noted that the non-monotonic behavior is mainly located in regions with $k_{x3} > \Lambda$, which means it mainly affects the unresolved small scales, and may or may not manifest itself in the resolved large scales. This non-monotonic behavior is observed and studied in \cite{Luo_2023}.

With these studies, it can be observed that including a small number of $k_x$ wavenumbers in the base flow using GQL is very effective at capturing the important triadic interactions for the forcing and even more effective for the response. However, the increase in $\Lambda$ does not guarantee a monotonic performance improvement of GQL, due to the changes in size and location of the neglected regions of triadic interactions. Finally, we will emphasize again that all the above analyses are performed using data from the DNS, a different dynamical system compared to QL/GQL. In QL/GQL, the modes will equilibrate at different amplitudes, shapes, and potentially phases due to the different dynamics compared to the DNS. Therefore, capturing triadic interactions shown to be important by the DNS data is not a sufficient condition, yet it is beneficial and likely a necessary condition for the success of reduced models.

\section{Summary and outlook}
In this work, we characterized spatio-temporal, resonant triadic interactions in turbulent channel flow, which arise due to the quadratic nonlinearity in the Navier-Stokes equations viewed from the Fourier domain. We proposed forcing and response coefficients to quantify the contribution from each pair of interacting wavenumber-frequency triplets to the resulting nonlinear forcing and velocity response. Building upon previous studies that focused on the transfer of energy between modes, we incorporated the linear resolvent operator into the response coefficients to provide the missing link from energy transfer into (or out of) a mode to the changes in the spectral turbulent kinetic energy of this mode. This provides a new and more complete description of the effect of triadic interactions on the resulting turbulent kinetic energy at each wavenumber-frequency triplet. 

At the level of individual scale triads, 
the important interactions are localized in temporal frequencies around a plane where the wavespeeds of all three participating scales are the same. This is mainly due to the quadratic nature of the nonlinear forcing and the spatial localization (in $y$) of Fourier modes associated with the critical layer mechanism in this space-time formulation. 
It was found that large forcing magnitude associated with an individual nonlinear interaction of scales did not necessarily correspond to a large velocity response. Although the forcing may have significant projection onto sub-optimal resolvent modes in a resolvent mode expansion, the velocity response is still dominated by the optimal resolvent response modes due to the linear amplification by the low-rank nature of the resolvent, consistent with only the solenoidal forcing exciting a velocity response. This again underscores the different perspectives offered by the inclusion of the linear resolvent operator into the analyses of the nonlinear triadic interactions. Note that the action of the dilatational forcing has not been examined here and remains a topic for future work.

The forcing and response coefficients calculated across scales highlight the importance of interactions involving large-scale structures, which is shown to be mainly driven by the interactions between large scales, while interactions between small scales are also non-negligible energy extraction mechanisms. For the small scales, it is revealed that the triadic interactions involving large scales contribute significantly, consistent with the coherence revealed by amplitude modulation studies. Finally, the phases of the coefficients are also utilized to reveal the constructive and destructive energy contributions by the triadic interactions. 

The importance of the large scales provides a natural connection to the modeling assumptions of the QL and GQL analyses. A detailed study of the regions of triadic interactions permitted under QL and GQL reductions revealed that they efficiently capture important triadic interactions in the flow, and the inclusion of small numbers of wavenumbers into the GQL large-scale base flow quickly captures most of the important triadic interactions. We emphasize that $\rho_f$, $\rho_r$ and $\gamma_f$, $\gamma_r$ are measurements of the contribution of the interactions permitted under QL/GQL reductions to the total forcing and response calculated by DNS of the full NSE. As such, it gives an indication of a possible reason for the success of QL and GQL simulations in replicating features of wall turbulence, without consideration of the different dynamics associated with the restricted systems. Additionally, a detailed analysis of regions of neglected triadic interactions in GQL is also performed. It is revealed that as $\Lambda$ increases, although more triadic interactions are included, certain interactions can still be lost. The relative importance between the lost and newly included triadic interactions can cause non-monotonic performance for the small scales, and may also have an effect on the resolved large scales in GQL. 


For future research, the methods presented here can be applied to more complex flows and could assist in the understanding of the underlying nonlinear mechanisms behind less understood flows. These tools can also be used with QL and GQL reductions to further quantify the underlying effect of the truncation of permitted interactions. Finally, it would be valuable to explore the potential modeling and computational benefits that can result from this sparsification of important triadic interactions within the flow.





\backsection[Funding]{The support of the U.S. Office of Naval Research under grant numbers N00014-17-1-2960 and N00014-17-1-3022 is gratefully acknowledged. The National Center for Atmospheric Research is sponsored by the National Science Foundation of the United States.}

\backsection[Declaration of interests]{The authors report no conflict of interest.}




\bibliographystyle{jfm}
\bibliography{./main}
\end{document}